\documentclass[reprint,onecolumn,superscriptaddress,amsmath,amssymb,aps,pre,longbibliography,nofootinbib]{revtex4-2}

\usepackage{enumitem}
\usepackage{amsmath}
\usepackage{amssymb}
\usepackage{mathtools}
\usepackage{graphicx}
\usepackage{bm}
\usepackage[dvipsnames]{xcolor}
\usepackage{accents}
\usepackage[colorlinks=true,allcolors=red!50!black]{hyperref}

\usepackage{tikz}
\usetikzlibrary{calc} 
\usetikzlibrary{decorations.pathmorphing}

\DeclareMathOperator{\tr}{tr}
\DeclareMathOperator{\Tr}{Tr}

\def\W{\bm W}
\def\w{\bm w}
\def\x{\bm x}
\def\g{\bm g}

\def\M{\bm M}

\def\G{\bm G}

\def\Sh{\Sh}

\def\T{\bm T}
\def\v{\bm v}

\def\y{\bm y}

\def\A{\bm A}
\def\B{\bm B}
\def\O{\bm O}
\def\F{\bm F}

\def\C{\bm C}
\def\X{\bm X}

\def\S{\bm \Sigma}

\def\df{\mathrm{df}}
\def\tf{\mathrm{tf}}
\def\Sh{\hat {\bm \Sigma}}

\def\Sig{\bm \Sigma}

\DeclareMathOperator*{\argmin}{arg\,min}

\usepackage{array}
\usepackage{booktabs}

\newcolumntype{L}[1]{>{\raggedright\arraybackslash}p{#1}}
\newcolumntype{M}[1]{>{\raggedright\arraybackslash}m{#1}}

\numberwithin{equation}{section}

\usepackage{autonum}

\begin{document}

\title{Two-point deterministic equivalence for stochastic gradient dynamics in linear models }

\author{Alexander Atanasov}
\altaffiliation{A.A., B.B., and J.A.Z.-V. contributed equally to this work.}
\email{atanasov@g.harvard.edu}
\affiliation{John A. Paulson School of Engineering and Applied Sciences, Harvard University, Cambridge, MA, USA}

\author{Blake Bordelon}
\altaffiliation{A.A., B.B., and J.A.Z.-V. contributed equally to this work.}
\email{blake\_bordelon@g.harvard.edu}
\affiliation{Center of Mathematical Sciences and Applications, Harvard University, Cambridge, MA, USA}

\author{Jacob A. Zavatone-Veth}
\altaffiliation{A.A., B.B., and J.A.Z.-V. contributed equally to this work.}
\email{jzavatoneveth@fas.harvard.edu}
\affiliation{Society of Fellows, Harvard University, Cambridge, MA, USA}
\affiliation{Center for Brain Science, Harvard University, Cambridge, MA, USA}

\author{\\Courtney Paquette}
\email{courtney.paquette@mcgill.ca}
\affiliation{Department of Mathematics and Statistics, McGill University, Montr{\'e}al, QC, CA}

\author{Cengiz Pehlevan}
\email{cpehlevan@seas.harvard.edu}
\affiliation{John A. Paulson School of Engineering and Applied Sciences, Harvard University, Cambridge, MA, USA}
\affiliation{Center for Brain Science, Harvard University, Cambridge, MA, USA}
\affiliation{Kempner Institute for the Study of Natural and Artificial Intelligence, Harvard University, Cambridge, MA, USA}

\begin{abstract}
We derive a novel deterministic equivalence for the two-point function of a random matrix resolvent. Using this result, we give a unified derivation of the performance of a wide variety of high-dimensional linear models trained with stochastic gradient descent. This includes high-dimensional linear regression, kernel regression, and linear random feature models. Our results include previously known asymptotics as well as novel ones. 
\end{abstract}

\maketitle

\section{Introduction}

Modern deep learning practice is governed by the surprising predictability of performance improvement with increases in the scale of data, model size, and compute \cite{hestness2017deep}. Often, the scaling of performance as a function of these quantities exhibits remarkably regular power law behavior, termed a neural scaling law \cite{kaplan2020scaling, hoffmann2022training,ghorbani2021scaling, hernandez2021scaling, hernandez2022scaling, gordon2021data, muennighoff2024scaling, zhai2022scaling, alabdulmohsin2024getting, bachmann2024scaling}. Here, performance is usually measured by some differentiable loss on the predictions of the model on a held out test set representative of the population. Given the relatively universal behavior of the exponents across architectures and optimizers \cite{kaplan2020scaling,hoffmann2022training,everett2024scaling}, one might hope that relatively simple models of information processing systems might be able to recover the same types of scaling laws. 

The (stochastic) gradient descent (SGD) dynamics in kernel methods \cite{lin2017optimal, pillaud2018statistical, ying2008online} and random feature models \cite{bordelon2024dynamical, paquette20244plus, lin2024scaling, carratino2018learning} were analyzed in recent works, exhibiting a surprising breadth of scaling behavior and capturing several interesting phenomena observed in deep network training. Each of the above works has isolated various effects that can hurt performance compared to the idealized infinite data and infinite model size limits. The model was first studied in \cite{bordelon2024dynamical}, where the bottlenecks due to finite width and finite dataset size were computed and, for certain data structure, resulted in a Chinchilla-type scaling result as in \cite{hoffmann2022training}. In \cite{paquette20244plus}, the effect of finite models size and online SGD noise was studied, and it was shown that under certain conditions these effects could lead to worse scaling exponents with the number of iterations than those one would naively calculate from a leading order picture. 

In this work, we aim to unify these prior results by providing a novel deterministic equivalence result for correlations of resolvent matrices evaluated at an arbitrary pair of arguments. This result allows us to combine the interaction of limited data, limited features, and SGD noise to model the stochastic process induced by SGD in the random feature model. We leverage the connections between deterministic equivalence and free probability highlighted in \cite{atanasov2024scaling}. Specifically, we use the properties of the $S$-transform and a planar diagrammatic expansion to obtain the asymptotic expressions for the train and test losses in time in linear random feature models. Our results recover those obtained via a dynamical mean field theory (DMFT) approach by \cite{bordelon2024dynamical} and with those obtained using deterministic equivalence techniques in \cite{paquette20244plus}. 

Concretely, the structure and contributions of the paper are as follows
\begin{itemize}
    \item In Section \ref{sec:setup}, we give the setup for the model that we consider, namely a linear random feature model of hidden width $N$ trained for $t$ steps with SGD of batch size $B$, at each step sampling with replacement from a training set of size $P$. We define the relevant objects that will arise in the study of its learning dynamics. 
    \item In Section \ref{sec:sgd}, we derive a reduced model for the dynamics of training and generalization error under SGD, following \cite{bordelon2022learning} and \cite{paquette20244plus}. We take the continuous time limit and highlight how random matrix quantities of interest naturally arise in Fourier space. We comment on when and how the large $t$ dynamics can be obtained from a Fourier space perspective.
    \item In Section \ref{sec:2_pt}, we derive a novel class of deterministic equivalences, which we term \textit{two-point} deterministic equivalences\footnote{Here, we take inspiration from the field theory language of ``$n$-point correlation functions'' \cite{zinn2021quantum}.}, since they involve two resolvents evaluated at different arguments, $\lambda$ and $\lambda'$. 
    \item In Section \ref{sec:LR}, we apply this to the simpler setting of a linear regression model trained with SGD on a given training set. We apply the equivalences derived to obtain sharp asymptotics for both the gradient flow term and the SGD kernel term.  We recover out-of-distribution results in prior literature and extend them to the setting of dynamics. 
    \item In Section \ref{sec:RF}, we extend this argument to the linear random feature model setting. Again, we derive sharp asymptotics for both the gradient flow and SGD kernel terms.
    \item Finally, in Section \ref{sec:DMFT} we comment on the relationship between the two-point deterministic equivalences and dynamical mean field theory, giving a new interpretation of the $S$-transform of free probability as a response function.
\end{itemize}

\section{Setup}\label{sec:setup}

In this paper, we will consider two types of models. The first will be given by a linear regression: 
\begin{equation}\label{eq:LR_dfn}
    f(\x) = \x^{\top} \w,
\end{equation}
where $\w$ is trainable. The second will be given by linear random features:
\begin{equation}\label{eq:LR_dfn}
    f(\x) = \x^{\top} \F \v,
\end{equation}
where $\v$ is trainable and $\F$ is fixed and random \cite{rahimi2007random}. Here, $\x \in \mathbb R^{D}$, $\F \in \mathbb R^{D \times N}$, $\v \in \mathbb R^{N}$. For the random feature model, we also define the \textit{effective learned weights} $\w \equiv \F \v$. In both cases, train $f(\x)$ to fit a set of labels $y$ generated from a noisy linear teacher $\bar \w$:
\begin{align}
    y_{\mu} = \bar{\w} \cdot \x_{\mu} + \epsilon_{\mu} .
\end{align}
We further have:
\begin{align}
    \x_\mu \sim \mathcal{N}(\bm{0}, \S), \quad \epsilon_\mu \sim \mathcal{N}(0, \sigma_{\epsilon}^2).
\end{align}
Here $\S$ is the covariance of the data, $\sigma_\epsilon^2$ is the variance of the label noise. Finally, we take $\F$ to have i.i.d. Gaussian entries with variance $1/N$. Note that this convention is different from that presented in \cite{atanasov2024scaling}, where $\F$ had entries of variance $1/D$. We choose this convention because it yields a more appropriate scaling of time.

We seek to minimize the empirical risk as a proxy for the population risk. Respectively, these are
\begin{align}
    \hat R = \frac{1}{P} \sum_{\mu=1}^{P} ( y_{\mu} - f(\x_{\mu}))^{2}, \quad R = \mathbb E_{\x, y} (y - f(\x))^2. %
\end{align}
Here, $P$ is the size of the finite training set that we have. We minimize this risk by running stochastic gradient descent (SGD). That is, at each step $t$, we sample a random batch $\mathcal B_t$ of $B < P$ training points (with replacement) and update the weights in proportion to the loss gradient on that batch: 
\begin{equation}
    \v_{t+1} = \v_{t} - \eta \nabla_{\v} \hat R_{\mathcal B_t} ,
\end{equation}
where $\eta > 0$ is the learning rate and
\begin{equation}
    \hat{R}_{\mathcal{B}_{t}} = \frac{1}{B} \sum_{\mu \in \mathcal{B}_{t}} ( y_{\mu} - f(\x_{\mu}))^{2}
\end{equation}
is the minibatch loss. 

As they will frequently appear in subsequent expressions, we define the design matrix $\X \in \mathbb R^{P \times D}$ and the vector of training labels $\y \in \mathbb R^{P}$. With this, we define the two empirical covariances:
\begin{equation}
    \Sh \equiv \frac{1}{P} \sum_{\mu = 1}^P \x_\mu \x_\mu^\top = \frac{1}{P} \X^\top \X , \quad  \Sh_t \equiv \frac{1}{B} \sum_{\mu \in \mathcal B_{t}} \x_{\mu,t} \x_{\mu,t}^\top.
\end{equation}
Note, $\mathbb E_{\mathcal B_t} \Sh_t = \Sh$ and $\mathbb E_{\X} \Sh = \S$. This highlights the two levels of randomness from the data: the randomness due to the choice of batch and the randomness due to the choice of the training set itself. It will be useful to appreciate that one can write  $\Sh = \S^{1/2} \W \S^{1/2}$ where $\W$ is an isotropic Wishart matrix (also known as a white Wishart matrix) with parameter $q = D/P$. See \cite{potters2020first, atanasov2024scaling} for details on the Wishart ensemble. Similarly, in our conventions, $\F^\top \F$ is a white Wishart matrix with parameter $D/N$.

An important point of comparison for our study will be the ridge estimator
\begin{align}
    \v_{\mathrm{Ridge}} &= \argmin_{\v} \bigg\{ \hat{R} + \lambda \Vert \v \Vert^2 \bigg\} 
    \\
    &= (\F^{\top} \Sh \F + \lambda )^{-1} \F^{\top} \frac{\X^{\top} \y}{P},
\end{align}
whose ``ridgeless'' ($\lambda \downarrow 0$) limit is equivalent to the infinite-time limit of full-batch gradient descent starting from zero initialization \cite{hastie2022surprises}. The behavior of ridge regression in high dimensions is well-understood, as reviewed in \cite{atanasov2024scaling}. 

\subsection{Degrees of Freedom}

In what follows, we will use $\tr$ to denote the normalized trace. For an $N \times N$ matrix $\A$, this is
\begin{equation}
    \tr[\A] = \frac1N \Tr[\A].
\end{equation}
The key quantities that emerge in the study of ridge regression in the case of statics are the \textit{degrees of freedom}. We define $\df^1_{\A}$ and $\df^2_{\A}$ as follows
\begin{equation}\label{eq:df_dfn}
    \df^1_{\A}(\lambda) \equiv \tr[\A (\A + \lambda)^{-1}], \quad
    \df^2_{\A}(\lambda) \equiv \tr[\A^2 (\A + \lambda)^{-2}].
\end{equation}
We will also write:
\begin{align}
    \df^2_{\A}(\lambda, \lambda') &\equiv \tr[\A^2 (\A + \lambda)^{-1} (\A + \lambda')^{-1}],
    \\
    \df^2_{\A , \A'}(\lambda, \lambda') &\equiv \tr[\A  \A' (\A + \lambda)^{-1} (\A + \lambda')^{-1}].
\end{align}
Notice the last definition is \textit{not} symmetric in $\A, \A'$. 
We also define the \textit{teacher-weighted} degrees of freedom as:
\begin{equation}
    \tf^1_{\A, \bar \w}(\lambda) \equiv \bar \w^\top \A (\A + \lambda)^{-1} \bar \w,
    \quad
    \tf^2_{\A, \bar \w}(\lambda) \equiv \bar \w^\top \A^2 (\A + \lambda)^{-2} \bar \w.
\end{equation}
These are defined as in \cite{atanasov2024scaling, atanasov2024risk,zavatone2023learning}.  

\subsection{$S$-transform}

Throughout this work, we will find that many key quantities can be expressed in terms of the $S$-transform of free probability. For a random matrix $\A$ drawn from some ensemble, we define the $S$-transform of the ensemble to be a function of the formal variable $\df$. Letting $\df_{\A}^{(-1)}(\df)$ be the functional inverse of $\df_{\A}^1$, namely $\df_{\A}^{(-1)}(\df_{\A}^1(\lambda) )= \lambda$, we have:
\begin{equation}\label{eq:S_defn}
    S_{\A}(\df) \equiv \frac{1-\df}{\df \,  \df_{\A}^{-1}(\df)}.
\end{equation}
To avoid ambiguity, we point out the denominator of the above equation merely a product of the variable $\df$ with $\df_{\A}^{-1}(\df)$. A consequence of this is that for all $\lambda$:
\begin{equation}\label{eq:S_defn2}
    \df^1_{\A}(\lambda) = \frac{1}{1+S_{\A}(\df^1_{\A}(\lambda)) \lambda}.
\end{equation}
The $S$-transform has the property that when two matrices $\A$, $\B$ are \textbf{free} of one another, one has the following property
\begin{equation}\label{eq:S_property}
    S_{\A * \B}(\df) = S_{\A}(\df) S_{\B} (\df).
\end{equation}
Here $\A * \B \equiv \A^{1/2} \B \A^{1/2}$ for $\A^{1/2}$ the principal matrix square root of $\A$. This is sometimes called the free product. A consequence of \eqref{eq:S_defn2} and \eqref{eq:S_property} is that for $\A, \B$ free of one another, one has:
\begin{equation}\label{eq:weak_det_equiv}
    \df^1_{\A  *\B} (\lambda) = \df^1_{\A} (\kappa)
\end{equation}
Here, $\kappa$ is known as the \textbf{resolution} or \textbf{signal capture threshold}, and can be calculated in two different ways. It can be calculated \textbf{empirically} as
\begin{equation}\label{eq:SB_def1}
   \kappa = \lambda S_{\B}(\df_{\A * \B}(\lambda)),
\end{equation}
or \textbf{omnisciently} via the self-consistent equation:
\begin{equation}\label{eq:SB_def2}
    \kappa = \lambda S_{\B}(\df_{\A}(\kappa)).
\end{equation}
Equation \eqref{eq:weak_det_equiv} is known as a \textbf{subordination relation} or as a  \textbf{weak deterministic equivalence}. Equations \eqref{eq:SB_def1} and \eqref{eq:SB_def2} are equivalent precisely because equation \eqref{eq:weak_det_equiv} holds.

The $S$-transform is particularly useful as it arises also in \textbf{strong deterministic equivalence}, where one can write
\begin{equation}
    \A \B ( \lambda + \bm A \bm B  )^{-1} \simeq \A (\A + \kappa)^{-1}.
\end{equation}
This is the un-traced form of the weak deterministic equivalence of equation \eqref{eq:weak_det_equiv}. Here, by $\M_1 \simeq \M_2$ we mean that $\tr[\M_1 \bm{\Theta}] / \tr[\M_2 \bm{\Theta}] \to 1$ as $N \to \infty$ for any sequence of test matrices $\bm{\Theta}$ of finite spectral norm \cite{bach2024high,atanasov2024scaling,potters2020first,knowles2017anisotropic,louart2018random,schroeder2024deterministic}.

In our linear random feature model setting, we have $D$ input dimensions, $N$ trainable parameters, and $P$ data points. We will take $D, N, P \to \infty$ jointly. In the high-dimensional case  where $\frac{1}{D} \Tr(\bm\Sigma) = \Theta_D(1)$ and each $\lambda_{i}(\bm\Sigma) = \Theta_{D}(1)$, taking $D, N, P \to \infty$ with fixed finite ratios yields a deterministic equivalence that is exact in the limit for training and test loss dynamics. More generally, $\Tr[\bm \Sigma]$ can be sublinear in $D$. For example, in the dimension-free setting $\Tr[\bm\Sigma]=\Theta_D(1)$. In all such cases, our planar diagrammatic analysis  computes the leading-order term in an asymptotic series in $D, N ,P$. Subleading terms (corresponding to non-planar diagrams) would contribute at finite $D, N, P$. These terms characterize both fluctuations of the risk about its mean value and corrections to the mean itself.

The relative accuracy of the leading order planar approximation depends on the spectral properties of $\bm \Sigma$ and the scaling relationships between $D, N, P$. However, based on known results in the static ridge regression setting \cite{defilippis2024dimension,misiakiewicz2024non}, and numerical results in previous work \cite{bordelon2024dynamical,atanasov2024scaling,paquette20244plus}, we conjecture that the asymptotic formulas we compute should nonetheless provide a good approximation to the risk. We will not attempt to prove quantitative error bounds in this work \cite{potters2020first,knowles2017anisotropic,louart2018random,schroeder2024deterministic,defilippis2024dimension,misiakiewicz2024non}. 

\section{Dynamics of SGD}\label{sec:sgd}

\subsection{Absorbing Effects of Label Noise}

Following the argument in \cite{canatar2021spectral}, one can straightforwardly account for the effect of label noise by allowing $\S$ to have an additional mode $\lambda_\infty$  going to zero and taking the corresponding mode in $\bar \w$, $\bar w_\infty$ to grow so that $\lambda_\infty \bar w_\infty = \sigma_\epsilon^2$ is fixed while $\lambda_\infty \to 0$. In the derivations that follow, we will assume the potential label noise has already been absorbed into appropriately defining $\bar \w, \S$, and not include it explicitly. Therefore, from here forward we will neglect the noise terms $\epsilon_\mu$, assuming they have been effectively absorbed into the definition of $\bar \w$ and $\S$. Note, as a result, that the label noise term will be included in the $\tf_1$ term. In particular, the following equation holds and will be used often:
\begin{equation}
    \lim_{\lambda \to 0} -\lambda^2 \tf_1'(\lambda) = \sigma_\epsilon^2.
\end{equation}

\subsection{A Reduced Model for SGD Risk Dynamics}

We now derive a reduced model for the dynamics of the training and generalization error under SGD, following previous works that argue this reduced model should be asymptotically equivalent in high dimensions \cite{paquette2021sgdlarge,paquette2022exactsgd,paquette2022implicit,paquette20244plus,bordelon2022learning}. We will not give a complete proof of this equivalence here, but instead reference conditions under which prior art gives either a rigorous proof or compelling evidence that it holds. 

In this analysis, we must make sure that we scale the learning rate $\eta$ and batch size $B$ correctly with the dimension $D$. For the batch size $B$, if $\bm{\Sigma}$ satisfies $\Tr(\bm\Sigma) = \Theta_D( D^{\zeta} )$ with $0 \leq \zeta \leq 1$, in which case the batch size $B$ should scale as $B = \Theta(D^{\zeta})$.  The learning rate $\eta$ is set independent of $D$. 

We first observe that all quantities of interest---namely the population loss $R_{t}$, the training loss $\hat R_t$, and the loss on a given batch $\hat R_{\mathcal B_t}$---can be written as quadratic forms in the weight discrepancy $\Delta \w_t \equiv \bar \w - \w_t$: 
\begin{equation}
    R_t = \Delta \w_t^\top \S \Delta \w_t, \quad \hat R_t = \Delta \w_t^\top \Sh \Delta \w_t, \quad \hat R_{\mathcal B_t} = \Delta \w_t^\top \Sh_t \Delta \w_t. 
\end{equation}
From the equations for SGD we have
\begin{equation}
\begin{aligned}
    \v_{t+1} &= \v_{t} - \eta \nabla_{\v} \hat R_{\mathcal B_t} =  v_t + \eta \F^\top \Sh_t \Delta \w_t\\
    \Rightarrow \Delta \w_{t+1} &= \Delta \w_{t} - \eta \F \F^\top  \underbrace{ \Sh_t \Delta \w_t}_{\g_t}.
\end{aligned}
\end{equation}

Using the fact that the covariates are Gaussian, we can compute the first two moments of $\g_t$:
\begin{equation}
    \mathbb E_{\mathcal B_t} \g_t =\Sh \Delta \w_{t},
\end{equation}
\begin{equation}
    \mathbb E_{\mathcal B_t} \g_t \g_t^\top = \Sh \Delta \w_{t} \Delta \w_{t}^\top \Sh + \frac{1}{B} \Sh \Delta \w_{t} \Delta \w_{t}^\top \Sh + \frac{1}{B} \Sh \underbrace{\Delta \w_{t}^\top \Sh \Delta \w_{t}}_{\hat R_t}.
\end{equation}
Because $R, \hat R$ are quadratic functions of $\Delta \w_{t}$, then following \cite{bordelon2022learning}, it is sufficient to track $\Delta \w_{t} \Delta \w_{t}^\top$:
\begin{equation}
\begin{split}
    \Delta \w_{t+1} \Delta \w_{t+1}^\top &= \Delta \w_{t} \Delta \w_t^\top - \eta \Delta \w_{t} \g_t \F \F^\top - \eta \F \F^\top \g_t \Delta \w_{t}^\top + \eta^2 \F \F^\top \g_t \g_t^\top \F \F^\top.
\end{split}
\end{equation}
Taking expectations over the batch and writing   $\C_t \equiv \mathbb E_{\mathcal B_t} \Delta \w_t \Delta \w_t^\top$ and $\chi = \eta /B$ for the \textbf{SGD temperature} yields: 
\begin{equation}\label{eq:full_dynamical_ODE}
\begin{split}
    \C_{t+1} &= (1- \eta \F \F^\top \Sh) \C_t (1- \eta \Sh \F \F^\top)  + \eta \chi \F \F^\top \Sh \C_t \Sh \F \F^\top  + \eta \chi  \F \F^\top \Sh \F \F^\top \Tr[\C_t \Sh]
\end{split}
\end{equation}
In what follows, we will drop the middle term in Equation \eqref{eq:full_dynamical_ODE}. We will now argue that this middle term is negligible for $\Tr( \bm\Sigma ) = \Theta(D^\zeta)$ with temperature scaling $\chi = \Theta(D^{-\zeta})$ for any $0 \leq \zeta \leq  1$. Depending on the structure of the data, this value either:
\begin{enumerate}
    \item explicitly vanishes in the high-dimensional $D \to \infty$ limit when $\zeta > 0$ 
    \item contributes negligibly after sufficient time $t$ for $\zeta = 0$.
\end{enumerate}
In the first case, maintaining stable dynamics with $\eta =\Theta(1)$ requires choosing a batch size large enough such that $\chi = \Theta( D^{-\zeta} )$ (see Section \ref{sec:kernel_terms}). With this choice, the two terms generated by SGD effects scale as
\begin{align}
    \eta \chi \F \F^\top \Sh \C_t \Sh \F \F^\top  = \Theta(D^{-\zeta}) \ , \  \eta \chi \F \F^\top \Sh \F \F^\top \Tr[\C_t \Sh] = \Theta(1) ,
\end{align}

justifying neglecting the left term. Alternatively, if the features are dimension free ($\zeta = 0$) so that $\Tr( \bm\Sigma ) < \infty$ as $D \to \infty$, then we note that the spectrum of $\bm\Sigma$ must decay sufficiently rapidly (the $k$-th ordered eigenvalue must decay faster than $1/k$). The projection of these terms along the $k$-th population eigendirection $\v_k$ are
\begin{align}
    \bm v_k^\top \F \F^\top \Sh \C_t \Sh \F \F^\top \bm v_k \approx \lambda_k^2 (\bar{w}_k)^2 e^{-2 \eta \lambda_k t}  \ , \   \bm v_k^\top \F \F^\top \Sh \F \F^\top   \bm v_k  \Tr[\C_t \Sh] \approx  \lambda_k \hat{\mathcal L}(t)
\end{align}
where $\mathcal{\hat L}(t) = \Tr[\C_t \Sh]$ is the training loss. For a small eigenvalue $\lambda_k \ll 1$ the first term will be dominated by the second term since it is quadratic in the small eigenvalue rather than linear. Further, for many features with $\Tr \bm\Sigma = \Theta(1)$, the training loss will decay more slowly than $e^{-\lambda_k t}$, leading to the final term dominating the SGD contribution to the update at large $t$.

Based on the two cases above, we thus consider the simplified dynamics
\begin{equation}\label{eq:associated_dynamical_ODE}
    \C_{t+1}= (1- \eta \F \F^\top \Sh) \C_t (1- \eta \Sh \F \F^\top)   + \eta \chi  \F \F^\top \Sh \F \F^\top \Tr[\C_t \Sh].
\end{equation}
We can now take the $\eta \to 0$ limit of this equation while keeping $\chi$ constant as in \cite{bordelon2022learning}. This will then include the SGD noise contributions to the training dynamics. By utilizing an integrating factor, we obtain the following differential equation:
\begin{equation}
    \frac{d}{dt} \left[ e^{\eta t \F \F^\top \Sh} \C_t  e^{\eta t \Sh \F \F^\top } \right] = \chi \, e^{2  t \F \F^\top \Sh}  \F \F^\top \Sh \F \F^\top  \Tr[\C_t \Sh].
\end{equation}
On the right-hand side we have applied the push-through identity to the matrix exponential. Integrating this, and using that $\C_0 = \bar \w \bar \w^\top$, we obtain the Volterra equation of \cite{paquette2021sgdlarge, paquette2022exactsgd, paquette20244plus}:
\begin{equation}\label{eq:Ct_evolution}
    \C_t \simeq e^{-\eta \F \F^\top \Sh} \bar \w \bar \w^\top e^{-\eta \Sh \F \F^\top } + \chi \int_0^t e^{-2  (t - s) \F \F^\top \Sh} \F \F^\top \Sh \F \F^\top \Tr[\C_s \Sh] ds
\end{equation}
Tracing against $\Sh$, $\S$ gives the evolution for the training and test losses respectively:
\begin{equation}\label{eq:volterra_emp_risk}
\boxed{
    \hat R_t = \underbrace{\bar \w^\top  e^{-2 t \Sh \F \F^\top } \Sh  \bar \w}_{ \hat {\mathcal F}(t)} \,  + \, \chi \int_0^t \underbrace{\tr[e^{-2 (t-s) \F \F^\top \Sh} (\F \F^\top \Sh)^2]}_{\hat {\mathcal K}(t-s)} \hat R_s ds,
    }
\end{equation}
\begin{equation}\label{eq:volterra_pop_risk}
\boxed{
\begin{aligned}
    R_t &= \underbrace{\bar \w^\top  e^{-t \Sh \F \F^\top } \S  e^{-t \F \F^\top \Sh} \bar \w}_{\mathcal F(t)} + \chi \int_0^t \underbrace{ \tr[e^{-2 (t-s) \F \F^\top \Sh} \F \F^\top \Sh \F \F^\top \S]}_{\mathcal K(t-s)} \hat R_s ds.
\end{aligned}}
\end{equation}
In both equations, the first term on the right hand side is referred to as the \textbf{forcing term}, while the second term we will refer to as the response or \textbf{kernel} term. It is the second term that is due to SGD noise, and would go away in the limit of $\chi \to 0$. The kernel term consists of a convolution of the population risk with the train and test kernels $\hat{\mathcal K}, \mathcal K$ respectively. This notation is adopted from \cite{paquette20244plus}. 

Importantly, in the case of nonzero $\sigma_\epsilon^2$, we see that $R_t$ will include a factor of the Bayes' error $\sigma_\epsilon^2$. We can also define the generalization error as the excess risk above this Bayes error:
\begin{equation}
    E_g \equiv R_t - \sigma_\epsilon^2.
\end{equation}
We now analyze both $R_t$ and $\hat R_t$ in more detail.

\subsection{Forcing Terms}

The forcing term in the equation for the generalization error will require a novel two-point deterministic equivalent to be derived. We first consider the more general quantity: 
\begin{equation}
    \mathcal F(t, t') = \Delta \w(t)^\top \Sig \Delta \w(t') = \bar \w^\top e^{- t \Sh \F \F^\top} 
    \S  e^{- t' \F \F^\top \Sh }  \bar \w.
\end{equation}
We want the diagonal $t = t'$ of this function. We first Fourier transform in $t, t'$ separately to obtain
\begin{equation}\label{eq:original_RF_forcing}
\begin{aligned}
    \mathcal F(t) &= \int_{\omega,\omega'} e^{i t (\omega + \omega')} \mathcal F(\omega, \omega')\\
    \mathcal F(\omega, \omega') &= \bar \w^\top (\Sh \F \F^\top + i \omega)^{-1} \S ( \F \F^\top \Sh + i \omega')^{-1} \bar \w.
\end{aligned}
\end{equation}
where we adopt the shorthand
\begin{align}
    \int_{\omega} (\cdot) = \frac{1}{2\pi} \int (\cdot) \,d\omega \quad \textrm{and} \quad \int_{\omega,\omega'} (\cdot) = \frac{1}{(2\pi)^2} \int (\cdot) \,d\omega\,d\omega'
\end{align}
for integrals over Fourier space. We see that in Fourier space, all of the randomness enters through the product of two resolvents evaluated at different ``imaginary ridges'', $i \omega, i \omega'$. The main technical goal of this note is to provide sharp asymptotics for this product, which we do in Section \ref{sec:2_pt}.

\subsection{Kernel Terms}\label{sec:kernel_terms}

Denoting temporal convolution by $\hat{ \mathcal K} \star \hat R$, we can rewrite Equation \eqref{eq:volterra_emp_risk} as
\begin{equation}
\begin{aligned}
    \hat R  &= \hat {\mathcal F} +\chi \hat{ \mathcal K} \star \hat R \\
    &= \hat {\mathcal F} + \chi \hat{ \mathcal K} \star \hat {\mathcal F} + \chi^2 \hat{ \mathcal K} \star \hat{ \mathcal K} \star \hat {\mathcal F} + \dots\\
    &= (1 - \chi \hat {\mathcal K})^{-1} \star \hat {\mathcal F}.
\end{aligned}
\end{equation}
Here, the last two lines are to be understood formally. For the test risk $R$ given in \eqref{eq:volterra_pop_risk}, an identical formal equation holds with $\hat{\mathcal{F}}$ replaced by $\mathcal{F}$. In order for SGD to be stable, we need a constraint that $\chi \|\mathcal K\| < 1$ in operator norm, such that this Neumann series converges. Examining the operator in Fourier space, 
\begin{align}
    \chi \mathcal K(\omega) =  \chi \  \Tr \  \left(\bm F \bm F^\top \hat{\bm\Sigma} \right)^2 [ i\omega  + \bm F \bm F^\top \hat{\bm\Sigma} ]^{-1} ,
\end{align}
it is clear that $\chi \Tr \bm\Sigma$ must be bounded for the dynamics to converge at late time (small $i\omega$) and for large data $P$ and large width $N$.

\subsection{Covariate Shift}

If the test set is out-of-distribution, in the sense that the covariates $\x$ are distributed according to a Gaussian with a different covariance $\S'$, the formalism can straightforwardly handle this, as shown in \cite{paquette2022implicit}. By tracing \eqref{eq:Ct_evolution} against the test covariance $\S'$, we have that the out-of-distribution generalization error $R_t'$ is given by:
\begin{equation}\label{eq:volterra_OOD_risk}
\begin{split}
    R_t^{OOD} &= \underbrace{\bar \w^\top  e^{-t \Sh \F \F^\top } \S'  e^{-t \F \F^\top \Sh} \bar \w}_{\mathcal F_{OOD}(t)}  + \chi \int_0^t \underbrace{ \tr[e^{-2 (t-s) \F \F^\top \Sh} \F \F^\top \Sh \F \F^\top \S']}_{\mathcal K_{OOD}(t-s)} \hat R_s ds.
\end{split}
\end{equation}

\subsection{Recovering Statics}\label{sec:final_value}

The long time limit of both the forcing function and the kernel function can be studied in Fourier space a well. In the single frequency setting, we require that all poles of the function in question lie in \textit{either} the upper half-plane or at $\omega = 0$ on the real line.  We see that the residues of any pole will be multiplied by the factor $e^{i \omega t}$. This goes to zero as $t \to \infty$ for any $\omega$ with $\Im( \omega ) > 0$. The remaining poles are on the real line. Poles away from zero would lead to oscillatory behavior at infinite time, and do not appear in the quantities that we treat. The same argument applies jointly to $\omega, \omega'$.

Assuming the remaining joint pole at $\omega, \omega' = 0$ is simple in both variables, we get that
\begin{align}
    \lim_{t \to \infty} \mathcal F(t) = \lim_{\omega, \omega' \to 0}  (i\omega) (i\omega') \mathcal F( i\omega, i\omega') .
\end{align}
This is known in the Laplace transform literature as the \textbf{final value theorem}, or more generally in physics as a \textbf{soft limit}.
The role of such soft limits in recovering static effects, also known as DC effects or ``memory effects'', has been highlighted in recent physics literature \cite{strominger2018lectures}.

\subsection{When is the One-Point Resolvent Insufficient?}

From the previous section, we see that in all of the settings of interest, the population forcing function (either in-distribution or out-of-distribution) can be represented as 
\begin{align}
    \mathcal F(t) = \bar{\w}^\top e^{- t \bm A \bm B } \bm M  e^{-t \bm B \bm A} \bar{\w}. 
\end{align}
where $\bm A$ and $\bm B$ are random symmetric matrices and $\bm M$ is a fixed symmetric matrix. Under the conditions that $\{\bm M, \bm A, \bm B \}$ all jointly commute, one can write this expression in Fourier space as:
\begin{align}
    \mathcal F(t) = \bar{\w}^\top e^{- 2 t \bm A \bm B } \bm M  \bar{\w}  = \int_{\omega} e^{2 i \omega t} \bar{\w}^\top \left(  i\omega + \bm A \bm B  \right)^{-1} \bm M \bar{\w}.
\end{align}
Under this commutativity condition, computing sharp asymptotics at finite time for the above quantity amounts to computing a \textbf{strong deterministic equivalence} for the random matrix $( i\omega + \bm A \bm B  )^{-1}$. Such strong deterministic equivalences have been highlighted in a variety of recent literature \cite{loureiro2021learning, bach2024high, atanasov2024scaling}. 
However, under a variety of settings, the matrices $\{\bm A, \bm B, \bm M\}$ will not commute, necessitating a different approach. Several such settings of interest are as follows:
\begin{enumerate}
    \item When the dataset is finite, $\A = \Sh$ and the population covariance $\S$ does not generally commute with $\A$, regardless of whether $\B$ is included. Thus, even in the case of linear regression at finite $t$, one requires two-point resolvents. 
    \item A random feature model at finite model size $N$ and finite dataset size $P$ will not have $\bm A = \bm F \bm F^\top $ and $\bm B = \hat{\bm \Sigma}$ commute. This is why the general finite $N, P, t$ expressions of Bordelon et. al \cite{bordelon2024dynamical} required computing two point resolvents that are functions of two frequencies $\omega, \omega'$. At finite $N$ but infinite $P$, $\A = \M = \S$ and one-point resolvents are in fact sufficient as in the work of Paquette et al \cite{paquette20244plus}.
    \item For covariate shift settings, the matrix $\bm M = \S'$ does not generally commute with $\bm A$ and $\bm B$. Even in the linear regression setting, where $\B = \mathbf I$, one still requires a two-point deterministic equivalent in this case \cite{patil2024ood,atanasov2024risk}.
\end{enumerate}
In cases 1. and 2. the $t \to \infty$ limits of the forcing term recover ridgeless regression, whose precise asymptotics does not require two-point equivalences, as demonstrated in \cite{atanasov2024scaling}. To address these more general settings, in the next section we derive a novel set of ``two-point'' deterministic equivalences.

\section{Two-Point Deterministic Equivalence}\label{sec:2_pt}

Let $\A, \M$ be deterministic and $\B$ be an isotropic multiplicative noise source.  All are $N \times N$ matrices and $\B$ is free of $\A$, $\M$.  We will eventually specialize the case in which $\B$ is a white Wishart matrix, but our derivations hold for general $\B$ satisfying the freeness assumption. We are interested in finding a deterministic equivalent for the following expression:
\begin{equation}\label{eq:to_eval}
    (\lambda + \A \B)^{-1} \M (\lambda' + \B \A)^{-1}.
\end{equation}
We call this a \textbf{two point resolvent}, by analogy to similar quantities in field theory that involve the insertion of an operator, in this case $(\lambda + \A \B)^{-1}$, at two different points, in this case $\lambda, \lambda'$.
We will use the following shorthand to simplify our final equations:
\begin{equation}
\begin{aligned}
    S_{\B} &= S_{\B}(-\df^1_{\A \B}(\lambda)), \quad & S_{\B}' &= S_{\B}(-\df^1_{\A \B}(\lambda')) ,\\
    \kappa &= \lambda S_{\B}, \quad & \kappa' &= \lambda' S_{\B}', \\
    \G_{\A} &= (\kappa + \A)^{-1}, \quad & \G_{\A}' &= (\kappa' + \A)^{-1}, \\
    g &= \tr[\G_{\A}], \quad & g' &= \tr[\G'_{\A}] ,\\
    \T_{\A} &=  \A (\kappa + \A)^{-1}, \quad & \T_{\A}' &= \A (\kappa' + \A)^{-1}. \\
\end{aligned}
\end{equation}
We first review a variation of the argument in \cite{atanasov2024scaling} to evaluate a single $\G_{\A}$. For more details on the orthogonal averages, the reader is encouraged to first go through Section III of that work. Following that argument, we utilize the freeness of $\B$ relative to $\A$ to write it as $\O \B' \O^\top$ with $\B'$ diagonal, and perform averages over the orthogonal matrix $\O$. In performing the average over the orthogonal group, using the language of \cite{atanasov2024scaling}, we can expand the resolvent in terms of a series of \textit{irreducible} diagrams linked together by multiplications by $\A/\lambda$. By orthogonal invariance, the irreducible diagrams must be scalars, equal to a value $1/S_{\B}$. This re-sums to $S_{\B} G_{\A}$. Diagrammatically, we write:
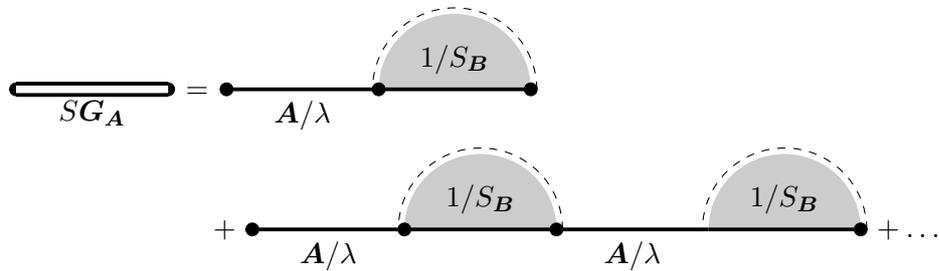
\begin{equation}
\begin{aligned}
\begin{tikzpicture}[baseline=-0.65ex, scale=0.8]
\def\spacing{2.5cm}
    \def\shift{0.1cm}
    \def\linewidth{0.5mm} %

     \foreach \x in {-1,0}
    \filldraw[black] (\x*\spacing,0) circle (3pt) node[below] {};

    \foreach \x in {-1}
    \draw[double, double distance=1mm, line width=\linewidth] (\x*\spacing,0) -- node[below] {$S \G_{\A}$} (\x*\spacing+\spacing,0);    
\end{tikzpicture}
 &= 
 \begin{tikzpicture}[baseline=-0.65ex, scale=0.8]
    \def\spacing{2.5cm}
    \def\shift{0.1cm}
    \def\linewidth{0.5mm} %

    \fill[gray!40] (1*\spacing,0) arc[start angle=0, end angle=180, radius=\spacing/2] -- cycle;
   
  \draw[dashed] (1*\spacing +  \shift,0 ) arc[start angle=0, end angle=180, radius=\spacing/2+\shift];
   
   \foreach \x in {-1}
    \draw[line width=\linewidth] (\x*\spacing,0) -- node[below] {${\A/\lambda}$} (\x*\spacing+\spacing,0);
    
   \node at ($(0*\spacing,0)!0.5!(1*\spacing,0)+(0,\spacing/5)$) {$1/S_{\B}$};
   
  \foreach \x in {-1,0,1}
    \filldraw[black] (\x*\spacing,0) circle (3pt) node[below] {};
  
  \foreach \x in {0}
    \draw[line width=\linewidth] (\x*\spacing,0) -- node[below] {} (\x*\spacing+\spacing,0);
\end{tikzpicture}\\
& \quad +  
\begin{tikzpicture}[baseline=-0.65ex, scale=0.8]
    \def\spacing{2.5cm}
    \def\shift{0.1cm}
    \def\linewidth{0.5mm} %

    \fill[gray!40] (1*\spacing,0) arc[start angle=0, end angle=180, radius=\spacing/2] -- cycle;

    \fill[gray!40] (3*\spacing,0) arc[start angle=0, end angle=180, radius=\spacing/2] -- cycle;
   
  \draw[dashed] (1*\spacing +  \shift,0 ) arc[start angle=0, end angle=180, radius=\spacing/2+\shift];

  \draw[dashed] (3*\spacing +  \shift,0 ) arc[start angle=0, end angle=180, radius=\spacing/2+\shift];
   
   \foreach \x in {-1,1}
    \draw[line width=\linewidth] (\x*\spacing,0) -- node[below] {${\A/\lambda}$} (\x*\spacing+\spacing,0);
    
   \node at ($(0*\spacing,0)!0.5!(1*\spacing,0)+(0,\spacing/5)$) {$1/S_{\B}$};

   \node at ($(4*\spacing,0)!0.5!(1*\spacing,0)+(0,\spacing/5)$) {$1/S_{\B}$};
   
  \foreach \x in {-1,0,1,3}
    \filldraw[black] (\x*\spacing,0) circle (3pt) node[below] {};
  
  \foreach \x in {0,2}
    \draw[line width=\linewidth] (\x*\spacing,0) -- node[below] {} (\x*\spacing+\spacing,0); 
\end{tikzpicture}
+ \dots
\end{aligned}
 \end{equation}
Here, again the language of \cite{atanasov2024scaling}, each irreducible diagram can be expressed as a sum of \textit{fully connected} diagrams:
 \begin{equation}
\begin{aligned}
    \begin{tikzpicture}[baseline=-0.65ex, scale=0.8]
    \def\spacing{2.5cm}
    \def\shift{0.1cm}
    \def\linewidth{0.5mm} %

    \fill[gray!40] (1*\spacing,0) arc[start angle=0, end angle=180, radius=\spacing/2] -- cycle;
   
  \draw[dashed] (1*\spacing +  \shift,0 ) arc[start angle=0, end angle=180, radius=\spacing/2+\shift];
    
   \node at ($(0*\spacing,0)!0.5!(1*\spacing,0)+(0,\spacing/5)$) {$1/S_{\B}$};
   
  \foreach \x in {0,1}
    \filldraw[black] (\x*\spacing,0) circle (3pt) node[below] {};
  
  \foreach \x in {0}
    \draw[line width=\linewidth] (\x*\spacing,0) -- node[below] {} (\x*\spacing+\spacing,0);
\end{tikzpicture} \quad \simeq \quad &
    \begin{tikzpicture}[baseline=-0.65ex, scale=0.6]
        \def\spacing{4cm}
        \def\shift{0.1cm}
        \def\linewidth{0.5mm}
        \foreach \x in {0}
            \filldraw[black] (\x*\spacing,0) circle (3pt) node[below] {$\B$};

        \draw[dashed] (0 ,0 ) arc[start angle=-90, end angle=270, radius=\spacing/4 ];
        
    \end{tikzpicture}
      +  
    \begin{tikzpicture}[baseline=-0.65ex, scale=0.8]
        \def\spacing{2.5cm}
        \def\shift{0.1cm}
        \def\linewidth{0.5mm} 
         \fill[gray!30] (1*\spacing +  \shift,0) arc[start angle=0, end angle=180, radius=\spacing/2 + \shift];
        
            \fill[white] (1*\spacing -  \shift,0) arc[start angle=0, end angle=180, radius=\spacing/2 - \shift];
                
        \foreach \x in {0,...,1}
            \filldraw[black] (\x*\spacing,0) circle (3pt) node[below] {$\B$};

        \foreach \x in {0}
            \draw[line width=\linewidth] (\x*\spacing,0) -- node[below] {$\bm G_{\A \B} \A$} (\x*\spacing+\spacing,0);

        \draw[dashed] (\spacing + \shift,0 ) arc[start angle=0, end angle=180, radius=\spacing/2 + \shift];
        \draw[dashed] (\spacing -  \shift,0 ) arc[start angle=0, end angle=180, radius=\spacing/2 - \shift];

    \end{tikzpicture}
\\&\quad +
    \begin{tikzpicture}[baseline=-0.65ex,scale=0.8]
        \def\spacing{2.5cm}
        \def\shift{0.1cm}
        \def
\linewidth{0.5mm} %

    \fill[gray!30] (2*\spacing + \shift,0) arc[start angle=0, end angle=180, radius=\spacing + \shift];

     \fill[white]  (2*\spacing -  \shift,0 ) arc[start angle=0, end angle=180, radius=\spacing/2 - \shift];
     \fill[white]  (\spacing -  \shift,0 ) arc[start angle=0, end angle=180, radius=\spacing/2 - \shift];
     
    \foreach \x in {0,...,1}
        \draw[line width=\linewidth] (\x*\spacing,0) -- node[below] {$\bm G_{\A \B} \A$} (\x*\spacing+\spacing,0);

    \draw[dashed] (2*\spacing + \shift,0) arc[start angle=0, end angle=180, radius=\spacing + \shift];
    \draw[dashed] (2*\spacing -  \shift,0 ) arc[start angle=0, end angle=180, radius=\spacing/2 - \shift];
    \draw[dashed] (\spacing -  \shift,0 ) arc[start angle=0, end angle=180, radius=\spacing/2 - \shift];

    \foreach \x in {0,...,2}
\filldraw[black] (\x*\spacing,0) circle (3pt) node[below] {$\B$};

\end{tikzpicture}  + \dots\\
&= \sum_{n=1}^\infty \kappa_{\B}^{(n)} \tr[\G_{\A \B} \A]^{n-1} =   \sum_{n=1}^\infty \kappa_{\B}^{(n)} (S \df_{\A})^{n-1} .
\end{aligned}
\end{equation}
Here, the shaded grey diagrams on the right hand side correspond to averages over the $\O$ that are, at leading order, identical to simple Wick contractions, but have subleading ``Weingarten'' terms that can still contribute to the final results. See section III of \cite{atanasov2024scaling} for a deeper discussion of this.
$\kappa_{\B}^{(n)}$ are the \textbf{free cumulants} of $\B$, as in \cite{potters2020first, atanasov2024scaling}. There, the $R$-transform is given by the power series in the formal variable $g$:
\begin{equation}
    R_{\B}(g) = \sum_{n=1}^\infty \kappa_{\B}^{(n)} g^{n-1}.
\end{equation}
Moreover, the $R$ and $S$ transform are related by the identity that $R_{\B}(S_{\B} \df) = S_{\B} (\df)^{-1}$. See \cite{potters2020first, atanasov2024scaling} for details. 
We thus have the strong deterministic equivalence from the prior work:
\begin{equation}
     G_{\A \B}(\lambda)  \simeq S_{\B} G_{\A}(\kappa), \quad S_{\B} = S_{\B}(\df_{\A}(\kappa)).
\end{equation}

Having reviewed this ``one-point'' deterministic equivalence of prior work, we now extend this approach to evaluate Equation \eqref{eq:to_eval}. We now expand in $1/\lambda$ and $1/\lambda'$ jointly. Before performing the orthogonal average, the general term will look like:
\begin{center}
\begin{tikzpicture}
    \def\spacing{2.5cm}
    \def\linewidth{0.5mm} %
  \foreach \x in {-1, 0, 1, 2}
    \filldraw[black] (\x*\spacing,0) circle (3pt) node[below] {$\O \B' \O^\top$};

  \foreach \x in {-2,-1}
    \draw[line width=\linewidth] (\x*\spacing,0) -- node[below] {$\A/\lambda$} (\x*\spacing+\spacing,0);
  \foreach \x in {0}
  \draw[line width=\linewidth] (\x*\spacing,0) -- node[below] {$\M$} (\x*\spacing+\spacing,0);
  \foreach \x in {1, 2}
    \draw[line width=\linewidth] (\x*\spacing,0) -- node[below] {$\A/\lambda'$} (\x*\spacing+\spacing,0);

  \node[below] at (-1*\spacing,0) {};
  \node[below] at (3*\spacing,0) {};
\end{tikzpicture}
\end{center}
Upon performing the average over $\O$, we recall that crossing diagrams do not contribute in the large $N$ limit. From this, we see that there will be two classes of terms in the diagrammatics. The first class is from averages that factorize into expectations over the individual resolvents $\G, \G'$. That is, the averages over the orthogonal group are performed separately to the left and to the right of $\M$. This ``disconnected'' contribution can be evaluated by appeal to the one-point equivalences separately on the left and right, and thus yields:
\begin{equation}
    \begin{tikzpicture}[baseline=-0.65ex, scale=0.8]
    \def\spacing{2.5cm}
    \def\shift{0.1cm}
    \def\linewidth{0.5mm} %
   
   \foreach \x in {-1}
    \draw[double, double distance=1mm, line width=\linewidth] (\x*\spacing,0) -- node[below] {$S_{\B} G_{\A}$} (\x*\spacing+\spacing,0);
    \foreach \x in {1}
    \draw[double, double distance=1mm, line width=\linewidth] (\x*\spacing,0) -- node[below] {$S_{\B}' G_{\A}'$} (\x*\spacing+\spacing,0);

  \foreach \x in {-1,0,1, 2}
    \filldraw[black] (\x*\spacing,0) circle (3pt) node[below] {};
 
   \foreach \x in {0}
    \draw[line width=\linewidth] (\x*\spacing,0) -- node[below] {$\M$} (\x*\spacing+\spacing,0);
  
\end{tikzpicture} \quad = \quad S_{\B} S'_{\B} \G_{\A} \M \G'_{\A} 
\end{equation}

The second term will involve averages that include a cumulant containing $\M$ underneath one of its arcs. Such terms will take the form:

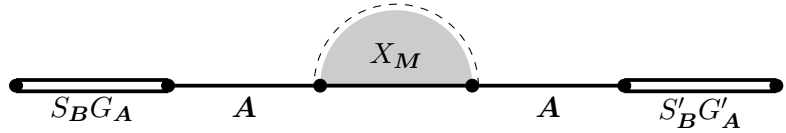
\begin{equation}
    \begin{tikzpicture}[baseline=-0.65ex, scale=0.8]
    \def\spacing{2.5cm}
    \def\shift{0.1cm}
    \def\linewidth{0.5mm} %
   
   \foreach \x in {-1}
    \draw[double, double distance=1mm, line width=\linewidth] (\x*\spacing,0) -- node[below] {$S_{\B} G_{\A}$} (\x*\spacing+\spacing,0);
    \foreach \x in {3}
    \draw[double, double distance=1mm, line width=\linewidth] (\x*\spacing,0) -- node[below] {$S_{\B}' G_{\A}'$} (\x*\spacing+\spacing,0);

    \foreach \x in {0, 2}
    \draw[line width=\linewidth] (\x*\spacing,0) -- node[below] {$\A$} (\x*\spacing+\spacing,0);

    \fill[gray!40] (2*\spacing,0) arc[start angle=0, end angle=180, radius=\spacing/2] -- cycle;
   
  \draw[dashed] (2*\spacing +  \shift,0 ) arc[start angle=0, end angle=180, radius=\spacing/2+\shift];
 
   \foreach \x in {1}
    \draw[line width=\linewidth] (\x*\spacing,0) -- node[below] {} (\x*\spacing+\spacing,0);

  \foreach \x in {-1,0,1, 2, 3, 4}
    \filldraw[black] (\x*\spacing,0) circle (3pt) node[below] {};

    \node at ($(1*\spacing,0)!0.5!(2*\spacing,0)+(0,\spacing/5)$) {$X_{\M}$};
  
\end{tikzpicture}
\end{equation}
Just like $S_{\B}$, $X_{\M}$ is again an expansion of fully connected diagrams, but under one of the arcs is an additional insertion of $\M$:
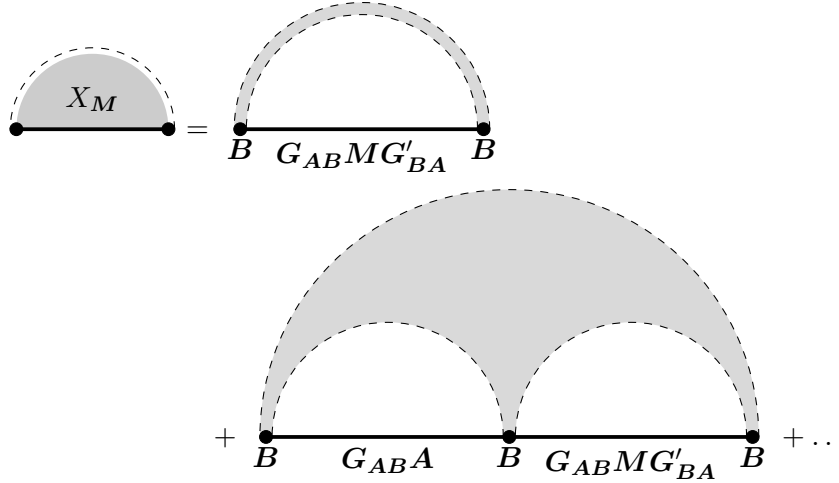
\begin{equation}
\begin{aligned}
        \begin{tikzpicture}[baseline=-0.65ex, scale=0.8]
    \def\spacing{2.5cm}
    \def\shift{0.1cm}
    \def\linewidth{0.5mm} %

    \fill[gray!40] (2*\spacing,0) arc[start angle=0, end angle=180, radius=\spacing/2] -- cycle;
   
  \draw[dashed] (2*\spacing +  \shift,0 ) arc[start angle=0, end angle=180, radius=\spacing/2+\shift];
 
   \foreach \x in {1}
    \draw[line width=\linewidth] (\x*\spacing,0) -- node[below] {} (\x*\spacing+\spacing,0);

  \foreach \x in {1, 2}
    \filldraw[black] (\x*\spacing,0) circle (3pt) node[below] {};

    \node at ($(1*\spacing,0)!0.5!(2*\spacing,0)+(0,\spacing/5)$) {$X_{\M}$};
\end{tikzpicture}
&=     \begin{tikzpicture}[baseline=-0.65ex, scale=0.8]
        \def\spacing{4cm}
        \def\shift{0.1cm}
        \def\linewidth{0.5mm} 
         \fill[gray!30] (1*\spacing +  \shift,0) arc[start angle=0, end angle=180, radius=\spacing/2 + \shift];
        
            \fill[white] (1*\spacing -  \shift,0) arc[start angle=0, end angle=180, radius=\spacing/2 - \shift];
                
        \foreach \x in {0,...,1}
            \filldraw[black] (\x*\spacing,0) circle (3pt) node[below] {$\B$};

        \foreach \x in {0}
            \draw[line width=\linewidth] (\x*\spacing,0) -- node[below] {$\bm G_{\A \B} \M \bm G_{\B \A}'  $} (\x*\spacing+\spacing,0);

        \draw[dashed] (\spacing + \shift,0 ) arc[start angle=0, end angle=180, radius=\spacing/2 + \shift];
        \draw[dashed] (\spacing -  \shift,0 ) arc[start angle=0, end angle=180, radius=\spacing/2 - \shift];

    \end{tikzpicture} \\&\quad +
        \begin{tikzpicture}[baseline=-0.65ex,scale=0.8]
        \def\spacing{4cm}
        \def\shift{0.1cm}
        \def
\linewidth{0.5mm} %

    \fill[gray!30] (2*\spacing + \shift,0) arc[start angle=0, end angle=180, radius=\spacing + \shift];

     \fill[white]  (2*\spacing -  \shift,0 ) arc[start angle=0, end angle=180, radius=\spacing/2 - \shift];
     \fill[white]  (\spacing -  \shift,0 ) arc[start angle=0, end angle=180, radius=\spacing/2 - \shift];
     
    \foreach \x in {0}
        \draw[line width=\linewidth] (\x*\spacing,0) -- node[below] {$\bm G_{\A \B} \A$} (\x*\spacing+\spacing,0);

    \foreach \x in {1}
        \draw[line width=\linewidth] (\x*\spacing,0) -- node[below] {$\bm G_{\A \B} \M \G'_{\B \A}$} (\x*\spacing+\spacing,0);

    \draw[dashed] (2*\spacing + \shift,0) arc[start angle=0, end angle=180, radius=\spacing + \shift];
    \draw[dashed] (2*\spacing -  \shift,0 ) arc[start angle=0, end angle=180, radius=\spacing/2 - \shift];
    \draw[dashed] (\spacing -  \shift,0 ) arc[start angle=0, end angle=180, radius=\spacing/2 - \shift];

\foreach \x in {0,...,2}
\filldraw[black] (\x*\spacing,0) circle (3pt) node[below] {$\B$};

\end{tikzpicture} + \dots 
    \end{aligned}
\end{equation}
Importantly, note that the factors of $1/\lambda$,  $1/\lambda'$ that do not accompany the $\A$ are indeed accounted for. They are absorbed into the $\G_{\A \B}$, $\G_{\A \B}'$ on both sides of $\M$.

In general, one can see that a given term in the expansion of $X_{\M}$ is labelled by two positive integers $a, b$. The first $a-1$ arcs will contain $\G_{\A \B} \A$ underneath, followed by a single arc containing $\G_{\A \B} \M \G'_{\B \A}$ underneath followed by $b-1$ arcs that will contain $\A \G'_{\B \A}$ underneath. This term will have $a+b$ insertions of $\B$ and thus will involve the $(a + b)$th cumulant of $\B$. \textit{Crucially}, as in the derivation of the $S$-transform, the $n$th cumulant only depend on the quantities under each arc \textit{through their traces}. We will denote the product of the first $a-1$ traces by $g^{a-1}$ and the product of the last $b-1$ traces by ${g'}^{b-1}$. One can  then write a self-consistent equation for $X_{\M}$ by recognizing the quantity under the arc with the $\M$ insertion is precisely the trace of the original two-point resolvent we sought to evaluate:
\begin{equation}
\begin{aligned}
    X_{\M} &=  S_{\B} S'_{\B} \sum_{n=1}^\infty \sum_{a+b=n} \kappa_{\B}^{(n)} g^{a-1} {g'}^{b-1}  \tr[G_{\A \B} \M G_{\A \B}]  \\
    &=  S_{\B} S'_{\B}  R_{\B}[g, g']  \Big( \tr[ \G_{\A} \M \G'_{\A} ] +  X_{\M} \tr[\G_{\A} \A^2 \G'_{\A}]  \Big).
\end{aligned}
\end{equation}
\begin{equation}
    R_{\B}[g, g'] = \sum_{n=1}^\infty \sum_{a+b=n} \kappa_{\B}^{(n)} g^{a-1} {g'}^{b-1}.
\end{equation}
where we have introduced the ``mixed'' $R$-transform
\begin{equation}
    R_{\B}[g, g'] = \sum_{n=1}^\infty \sum_{a+b=n} \kappa_{\B}^{(n)} g^{a-1} {g'}^{b-1}.
\end{equation}
We can then solve this self-consistent equation for $X_{\M}$: 
\begin{align}
    X_{\M} = \frac{S_{\B} S'_{\B} R_{\B}[g, g'] \tr[ \G_{\A} \M \G'_{\A} ] }{1 - S_{\B} S'_{\B} R_{\B}[g, g'] \tr[\G_{\A} \A^2 \G'_{\A}] } .
\end{align}
Therefore, we have the general deterministic equivalence
\begin{align}
    (\lambda + \A \B)^{-1} \M (\lambda' + \B \A)^{-1} \simeq  S_{\B} S'_{\B} \left[ \G_{\A} \M \G'_{\A} + X_{M} \G_{\A} \A^2 \G'_{\A}  \right]
\end{align}
with $X_{\M}$ as above. 

The ``mixed $R$-transform'' $R_{\B}[g,g']$ simplifies in the case when $\B$ is a white Wishart matrix $\B = \frac{1}{P} \X^\top \X$. There, because of the factorization property $\kappa_{\B}^{(a+b)} = q \kappa_{\B}^{(a)} \kappa_{\B}^{(b)}$, one has $R_{\B} [g, g'] = q  R_{\B}[g] R_{\B}[g'] = q (S_{\B} S'_{\B})^{-1}$ with $q = N/P$, so that
\begin{equation}
\begin{aligned}
X_{\M} =  \frac{q \tr[\G_{\A} \M \G_{\A}']}{1 - q \df_2(\kappa, \kappa')}, \quad \df_2(\kappa, \kappa') \equiv \tr[\A^2 \G_{\A} \G'_{\A}] .
\end{aligned}
\end{equation}
All together we get the final deterministic equivalence in the Wishart case:
\begin{equation}\label{eq:master_2pt_0}
\boxed{
\begin{aligned}
    (\lambda + \A \B)^{-1} \M (\lambda' + \B \A)^{-1} \simeq  S_{\B} S'_{\B} \left[ \G_{\A} \M \G'_{\A} +  \G_{\A} \A^2 \G'_{\A} \frac{q \tr[\G_{\A} \M \G_{\A}']}{1 - q \df_2(\kappa, \kappa')} \right].
\end{aligned} }
\end{equation}
The special case of this equivalent for $\lambda = \lambda'$ has appeared in our previous work \cite{atanasov2024risk} (there, we worked directly in the Wishart case, for which the diagrammatics directly represent Wick contractions). For Gram Wishart matrices, e.g. $\B = \frac{1}{P} \X \X^\top $ or $\B = \frac{1}{D} \F \F^\top$, the above equations hold but with $q \to 1/q$, since there $\kappa_{a+b} = q^{-1} \kappa_{a} \kappa_{b}$. We state a variety of additional variants of the above formula in Appendix \ref{app:all_two_point}. 

\section{Application I: Linear Regression}\label{sec:LR}

As a warm up, we first consider linear regression without random features. In this setting, one neglects $\F \F^\top$ by replacing it with the identity matrix. The dynamics in this setting are then:
\begin{equation}\label{eq:volterra_emp_risk_LR}
    \hat R_t = \underbrace{\bar \w^\top  e^{-2 t \Sh} \Sh  \bar \w}_{ \hat {\mathcal F}(t)} \,  + \, \chi \int_0^t \underbrace{\tr[e^{-2 (t-s) \Sh} \Sh^2]}_{\hat {\mathcal K}(t-s)} \hat R_s ds,
\end{equation}
\begin{equation}\label{eq:volterra_pop_risk_LR}
    R_t = \underbrace{\bar \w^\top  e^{-t \Sh } \S  e^{-t \Sh} \bar \w}_{\mathcal F(t)} \, + \, \chi \int_0^t \underbrace{ \tr[e^{-2 (t-s)  \Sh}  \Sh  \S]}_{\mathcal K(t-s)} \hat R_s ds.
\end{equation}

\subsection{Gradient Flow Term}

The generalization error in Fourier space can  then directly be obtained via the two-point master equation \eqref{eq:master_2pt_1} with $\M = \A = \S$ to obtain
\begin{equation}\label{eq:LR_population_forcing_det_equiv}
     \mathcal F(\omega, \omega')  =  \frac{S S'}{1-\gamma(i \omega_1, i \omega_1')} \bar \w^\top (i \omega_1 + \S)^{-1} \S (i \omega_1' + \S)^{-1} \bar \w .
\end{equation}
Here,  because $\Sh = \S * \W$ for $\W$ a white Wishart, the renormalization of each frequency is given by the multiplication of the $S$-transform of a white Wishart. This is found in the standard literature, see e.g. \cite{potters2020first}. Our notation convention in this section is thus:
\begin{equation}
\begin{aligned}
    \df_1 \equiv \df_{\Sh}^1(\omega) \simeq \df_{\S}^1(\omega_1),& \quad \df_1' \equiv \df_{\Sh}^1(\omega') \simeq \df_{\S}^1(\omega_1')\\
    S \equiv  \frac{1}{1 - \frac{D}{P} \df_1}, & \quad  S' \equiv \frac{1}{1 - \frac{D}{P} \df_1'}, \\
    \omega_1 \equiv S \omega, & \quad  \omega_1' \equiv S' \omega', \\
    \df_2 \equiv  \tr  \Big[ \S^2   (i \omega_1 & + \S)^{-1} (i \omega_1'  + \S)^{-1} \Big].\\
    \gamma \equiv & \frac{D}{P}  \df_2.
\end{aligned}
\end{equation}

The empirical forcing term can be handled with a single frequency Fourier transform. Indeed it is much more convenient to do so when dealing with the SGD effects of the next section. We can write it as:
\begin{equation}
    \hat{\mathcal F}(t) = \int_{\omega} e^{2 i \omega t} \bar \w^\top \Sh (\Sh + i \omega)^{-1} \simeq \int_{\omega} e^{2 i \omega t} \bar \w^\top \S (\S + i \omega_1)^{-1}.
\end{equation}
In the last equality, we have applied a strong (one point) deterministic equivalence.

We consider instead the bi-frequency transformation of the empirical loss in Appendix \ref{app:double_frequency_forcing}, and see that it gives a dynamical analogue of the generalized cross-validation (GCV) in the gradient flow limit. 

\subsection{SGD Kernel Term}

In order to correctly treat the convolution of the kernel term with the empirical risk, it is much easier to work in single frequency Fourier space. We now evaluate both the train and the test kernel.
\begin{equation}
    \mathcal K_t = \Tr[e^{- 2 t \Sh} \Sh \S] = \int_{\omega} e^{2 i \omega t}  \Tr[\S \Sh (\Sh + i \omega)^{-1}] .
\end{equation}
We can apply one-point deterministic equivalent to obtain:
\begin{equation}
      \Tr[\S \Sh (\Sh + i \omega)^{-1}]  \simeq \Tr[\S^2 (\S + i \omega_1)^{-1} ] = \Tr[\S] - i \omega_1 \Tr[\S (\S + i \omega_1)^{-1}]
\end{equation}
Similarly, for the train kernel, we have
\begin{equation}
    \hat{\mathcal K}_t = \Tr[e^{- 2 t \Sh} \Sh^2] = \int_{\omega} e^{2 i \omega t}  \Tr[\Sh^2 (\Sh + i \omega)^{-1}].
\end{equation}
In Fourier space, this resolvent is given by:
\begin{equation}
\begin{aligned}
    \Tr[\Sh^2 (\Sh + i \omega)^{-1}] &= \Tr[\Sh] - i \omega \Tr[\Sh (\Sh + i \omega)^{-1}] \\
    &\simeq \Tr[\S] - i \omega \Tr[\S (\S + i \omega_1)^{-1}]\\
    &= \Tr[\S^2 (\S+ i \omega_1)^{-1}]+i (\omega_1 - \omega) \Tr[\S (\S + i \omega_1)^{-1}].\\
    &= \mathcal K(\omega) + \frac{i \omega_1}{P} \Tr[\S (\S + i \omega_1)^{-1}]^2
\end{aligned}
\end{equation}
Where we have used that $i (\omega_1 - \omega) = i \omega (S -1) = i \omega_1 \frac{D}{P} \df_1$.
We thus see that the deterministic equivalent for the train kernel is the same as the equivalent for the test kernel, but with an extra additive term. In the large $t$, or equivalently $\omega \to 0$ limit, we have that this additional term is always \textit{non-negative}. 

Given the exact expression for the test risk, we get that the additive SGD contributions to the train and test risk are given by
\begin{equation}
\begin{aligned}
    \hat R(t) &= \int_{\omega} e^{2 i \omega t} \frac{\hat{\mathcal F}(\omega)}{1 - \hat{\mathcal K}(\omega)} 
    \\&= \int_{\omega} e^{2 i \omega t} \frac{\bar \w^\top \S (\S + i \omega_1)^{-1} \bar \w}{1 -  \chi \Tr[\S^2 (\S + i \omega_1)^{-1}] - \chi i \omega_1 (D \df_1)^2/P} 
\end{aligned}
\end{equation}
and
\begin{equation}
\begin{aligned}
    R(t) &= \int_{\omega,\omega'}  e^{i(\omega + \omega') t} \mathcal F(\omega, \omega') + \int_{\omega} e^{2 i \omega t}  \frac{\mathcal K(\omega) \hat{\mathcal F}(\omega)}{1-\hat{\mathcal K}(\omega)} \\
    &= \int_{\omega,\omega'}  \frac{  e^{i(\omega + \omega') t} S S'}{1-\gamma(i \omega_1, i \omega_1')} \bar \w^\top (i \omega_1 + \S)^{-1} \S (i \omega_1' + \S)^{-1} \bar \w \\
    & \quad + \int_{\omega} e^{2 i \omega t}  \frac{\chi \Tr[\S^2 (\S + i \omega_1)^{-1}]}{1 - \chi \Tr[\S^2 (\S + i \omega_1)^{-1}] - \chi i \omega_1 (D \df_1)^2/P } \bar \w^\top \S (\S + i \omega_1)^{-1} \bar \w   ,
\end{aligned}
\end{equation}
respectively. In Appendix \ref{app:double_frequency_kernel}, we compute the SGD terms in terms of double frequency Fourier transforms and two-point equivalents.

\subsection{Covariate Shift}

In this section, we show how the two-point deterministic equivalences allow for direct calculation of the train and test risks under a shift of covariates from $\S$ to $\S'$. There, the forcing term changes to
\begin{equation}\label{eq:LR_OOD}
\begin{aligned}
     \mathcal F_{OOD}(\omega, \omega') &=  S S' \bar \w^\top (i \omega_1 + \S)^{-1} \S' (i \omega_1' + \S)^{-1} \bar \w  + S S' \frac{ \gamma_{\S'} }{1-\gamma} \bar \w^\top (i \omega_1 + \S)^{-1} \S (i \omega_1' + \S)^{-1} \bar \w .
\end{aligned}
\end{equation}
In the above equation, 
\begin{equation}
    \gamma_{\S'} (i \omega_1, i \omega_1') \equiv \frac{D}{P} \tr \Big[ \S  (i \omega_1 + \S)^{-1} \S' (i \omega_1'  + \S)^{-1} \Big].
\end{equation}
The (single frequency) SGD test kernel term also changes to:
\begin{equation}
\begin{aligned}
    \mathcal K'(\omega) &\simeq \Tr[\S \S' (\S + i \omega_1)^{-1} ] = \Tr[\S'] - i \omega_1 D \df_1.
\end{aligned}
\end{equation}
In the static limit, we have $\kappa = \lim_{\omega=0} i \omega_1$. We also have explicitly written out the noise term rather than including it as a mode at infinity. Then,  the distribution-shifted gradient flow term becomes:
\begin{equation}
\begin{split}
    E_{OOD}^{\S', \bar \w} &\simeq \kappa^2 \left[ \bar \w^\top (\S + \kappa)^{-1} \S' (\S + \kappa)^{-1} \bar \w + \bar \w^\top \S (\S + \kappa)^{-2} \bar \w \frac{\gamma'}{1 - \gamma} \right] + \sigma_\epsilon^2 \frac{\gamma'}{1-\gamma}
\end{split}
\end{equation}
This recovers the covariate-shifted results of \cite{canatar2021out,atanasov2024risk,patil2024ood}. We note that one can also easily accommodate target shifts (\emph{i.e.}, from $\bar\w$ to $\bar\w'$) in this formalism, but doing so does not require two-point equivalents \cite{patil2024ood}.

\section{Application II: Random Feature Regression}\label{sec:RF}

We now return to our original model, with random features added. When data averaging, $\omega$ will be renormalized by the $S$-transform of a white Wishart $S_{\W}$ as before. When averaging over features, it will further be renormalized by the $S$-transform $S_{\F \F^\top}$. Here, $\F \F^\top$ is a white Wishart matrix with parameter $q=D/N$.  Our notation will thus be:
\begin{equation}
\begin{aligned}
    \df_1 \equiv \df_{\F \F^\top \Sh}^1(\omega), &\quad \df_1' \equiv \df_{\F \F^\top \Sh}^1(\omega') ,\\
    S_{\W} \equiv  \frac{1}{1 - \frac{D}{P} \df_1}, & \quad  S_{\W}' \equiv \frac{1}{1 - \frac{D}{P} \df_1'}, \\
    S_{\F\F^\top} \equiv  \frac{1}{1 - \frac{D}{N} \df_1}, & \quad  S_{\F\F^\top}' \equiv \frac{1}{1 - \frac{D}{N}\df_1'}, \\
    S \equiv S_{\W} S_{\F \F^\top}, & \quad S' \equiv S'_{\W} S'_{\F \F^\top},\\
    \omega_1 \equiv S_{\W} \omega, & \quad  \omega_1' \equiv S_{\W}' \omega', \\
    \omega_2 \equiv S_{\F \F^\top} \omega_1 = S\omega, & \quad  \omega_2'  \equiv S_{\F \F^\top}' \omega_1' = S' \omega',\\
    \df_2 \equiv \tr[\S^2 (\S +& i  \omega_2)^{-1} (\S + i \omega_2')^{-1}] .
\end{aligned}
\end{equation}
Then, the weak deterministic equivalents for $\df_{1}$ and $\df_{1}'$ can be written as
\begin{equation}
\begin{aligned}
    \df_1 \equiv \df_{\F \F^\top \Sh}^1(\omega) \simeq  \df_{\F \F^\top \S}^1(\omega_1) \simeq \df_{\S}^1(\omega_2), \\ \df_1' \equiv \df_{\F \F^\top \Sh}^1(\omega')  \simeq  \df_{\F\F^\top \S}^1(\omega_1') \simeq \df_{\S}^1(\omega_2') ,
\end{aligned}
\end{equation}
where we average first over the randomness in $\Sh$ and then $\F\F^{\top}$. As stated before, in contrast to \cite{atanasov2024scaling}, we choose to divide by $N$ instead of $D$ in defining the features. This will lead to a more natural gradient flow dynamics. In the convention of \cite{atanasov2024scaling}, one would have to re-scale $\eta \to \eta D /N$ to get correct dynamics. Either way, the large time limit will agree with agree with the (ridgeless) random feature results quoted in that work.

We will also make frequent use of the \textbf{push-through identity}:
\begin{equation}\label{eq:push_through}
    \A (\B \A + \lambda)^{-1} = (\A \B + \lambda)^{-1} \A.
\end{equation}

\subsection{Gradient Flow Term}

By writing $\Sh = \S^{1/2} \W \S^{1/2}$ for $\W$ a white Wishart matrix and applying the push-through identity  \eqref{eq:push_through} to the original random feature forcing equation \eqref{eq:original_RF_forcing}, we see that we need to evaluate:
\begin{equation}
\begin{aligned}
    \mathcal{F}(i \omega, i \omega') = \bar \w^\top \S^{1/2} & (i \omega + \W \S^{1/2}  \F \F^\top \S^{1/2})^{-1}  (i \omega' +    \S^{1/2} \F \F^\top \S^{1/2} \W )^{-1} \,  \S^{1/2} \bar \w.
\end{aligned}
\end{equation}
We do this in two steps, first integrating over data $\W$ and then over random features $\F \F^\top$.

\subsubsection{Integrating Over Data} Pushing through $\S^{1/2}$ on both sides, we apply the two-point equivalence \eqref{eq:master_2pt_3} with $\M = \mathbf I$ and $\A = \S^{1/2}  \F \F^\top \S^{1/2} $ to obtain: 
\begin{equation}
    \mathcal F(i \omega, i \omega') \simeq \frac{ S_{\W} S_{\W}'}{1-\gamma_1} \bar \w^\top (i \omega_1 + \S \F \F^\top )^{-1} \S  (i \omega_1' + \F \F^\top \S )^{-1} \bar \w
\end{equation}
for
\begin{equation}
    \gamma_1 \equiv \frac{D}{P} \tr\left[ ( \S \F \F^\top )^{2} (i \omega_1 + \S \F \F^\top)^{-1} (i \omega_1' +  \S \F \F^\top)^{-1} \right].
\end{equation}

\subsubsection{Integrating over Features}

We will evaluate $\gamma_1$ separately since it concentrates. We apply Equation \eqref{eq:master_2pt_2}  for a \textit{Gram Wishart} with $q = D/N, q^{-1} = N/D$. Here,  $\M=\mathbf I$, $\A = \S$, $B = \F \F^\top$.
\begin{equation}
\begin{aligned}
        \gamma_1 &\simeq \frac{D}{P} \df_2 + \frac{D}{P} \frac{D}{N} (i \omega_2)(i \omega'_2) \frac{   \tr[ (i \omega_2 + \S)^{-1} \S (i \omega'_2 +  \S)^{-1} ]^2 }{1 - \frac{D}{N} \df_2 }.
\end{aligned}
\end{equation}
We now apply  \eqref{eq:master_2pt_0} with $\M = \mathbf I$, $\A = \S$, and $\B = \F \F^\top$ to get:
\begin{equation}\label{eq:final_RF_forcing_term}
\begin{aligned}
     \mathcal F(i \omega, i \omega') &\simeq \frac{S S'}{1-\gamma_1} \Bigg[ \bar \w^\top (i \omega_2 + \S)^{-1} \S (i \omega'_2 +  \S)^{-1} \bar \w \\
     & \qquad  \qquad  \quad + \bar \w^\top (i \omega_2 + \S)^{-1} \S^2 (i \omega'_2 +  \S)^{-1} \bar \w  \frac{\frac{D}{N} \tr[ (i \omega_2 + \S)^{-1} \S (i \omega'_2 +  \S)^{-1}]}{1 - \frac{D}{N} \df_2(i \omega_2, i \omega_2')} \Bigg].
\end{aligned}
\end{equation}
This recovers the result of \cite{bordelon2024dynamical}. The formulas can also straightforwardly be extended to the case of covariate shift in the test set (Appendix \ref{app:RF_OOD}).

We now evaluate the training loss. Because of the push-through identity, even in the random feature setting we can characterize this in single-frequency space using a one-point deterministic equivalent:
\begin{equation}
\begin{aligned}
    \hat{\mathcal F}(t) &= \bar \w^\top e^{-t \Sh \F \F^\top} \Sh e^{-t \F \F^\top \Sh} \bar \w 
    =  \bar \w^\top \Sh   e^{-2t \F \F^\top \Sh } \bar \w\\
    &= \int_{\omega} e^{2 i \omega t} \bar \w^\top \Sh (\F \F^\top \Sh + i \omega)^{-1} \bar \w 
    \\
    &\simeq \int_{\omega} e^{2 i \omega t} S_{\F \F^\top} \bar \w^\top \S (\S + i \omega_2)^{-1} \bar \w .\\
\end{aligned}
\end{equation}
In the last line we have applied the one-point deterministic equivalence twice, over $\F \F^\top$ and over $\W$.

\subsection{SGD Kernel Term}

Here, we again apply a one-point deterministic equivalence to characterize the SGD kernel in single-frequency space. For the test kernel, we have:
\begin{equation}
\begin{aligned}
    \mathcal K(t) &= \int_{\omega} e^{2 i \omega t} \Tr[\S \F \F^\top \Sh \F \F^\top (\Sh \F \F^\top + i \omega)^{-1}]
\end{aligned}
\end{equation}
We apply the following two deterministic equivalences:
\begin{equation}
\begin{aligned}
    &\Tr[\S \F \F^\top \Sh \F \F^\top (\Sh \F \F^\top + i \omega)^{-1}]  \\
    &\quad\simeq \Tr[\S \F \F^\top \S \F \F^\top (\S \F \F^\top + i \omega_1)^{-1}] \\
    &\quad= \Tr[\S \F \F^\top] - i \omega_1 \Tr[\S \F \F^\top (\S \F \F^\top + i \omega_1)^{-1}]\\
    &\quad\simeq \Tr[\S] - i \omega_1 \Tr[\S (\S + i \omega_2)^{-1}].\\
    &\quad\simeq \Tr[\S^2 (\S + i \omega_2)^{-1}] + i (\omega_2 - \omega_1) \Tr[\S (\S + i \omega_2)^{-1}]\\
    &\quad\simeq \Tr[\S^2 (\S + i \omega_2)^{-1}] + \frac{i \omega_2}{N} \Tr[\S (\S + i \omega_2)^{-1}]^2.
\end{aligned}
\end{equation}
Similarly, for the train kernel, we have:
\begin{equation}
\begin{aligned}
    \hat{\mathcal K}(t) &= \int_{\omega} e^{2 i \omega t} \Tr[\Sh \F \F^\top \Sh \F \F^\top (\Sh \F \F^\top + i \omega)^{-1}] .
\end{aligned}
\end{equation}
Here, we apply two deterministic equivalences again:
\begin{equation}
\begin{aligned}
    &\Tr[\Sh \F \F^\top \Sh \F \F^\top (\Sh \F \F^\top + i \omega)^{-1}] \\
    &\quad= \Tr[\Sh \F \F^\top] - i \omega \Tr[\S \F \F^\top (\S \F \F^\top + i \omega)^{-1}] \\
    &\quad\simeq \Tr[\S ] - i \omega \Tr[\S (\S + i \omega_2)^{-1}].\\
    &\quad= \Tr[\S^2 (\S + i \omega_2)^{-1}] + i (\omega_2 - \omega) \Tr[\S (\S + i \omega_2)^{-1}]\\
    &\quad= \mathcal K(\omega) + i (\omega_1 - \omega) \Tr[\S (\S + i \omega_2)^{-1}]\\
    &\quad= \mathcal K(\omega)  + \frac{i \omega_1}{P} \Tr[\S (\S + i \omega_2)^{-1}]^2.
\end{aligned}
\end{equation}
All together, the empirical and population risk are then
\begin{equation}
\begin{aligned}
\hat R(t) &= \int_{\omega} e^{2 i \omega t} \frac{S_{\F \F^\top} \bar \w^\top \S (\S + i \omega_2)^{-1} \bar \w }{1 - \chi \Tr[\S] + i \omega  \chi D \df_1} ,\\
R(t) &= \int_{\omega,\omega'} e^{i(\omega + \omega')t} \mathcal F(\omega, \omega')  + \int_{\omega} e^{2 i \omega t}   \frac{\chi \Tr[\S] - i \omega_1 \chi D \df_1 }{1 - \chi \Tr[\S] + i \omega  \chi D \df_1} S_{\F \F^\top} \bar \w^\top \S (\S + i \omega_2)^{-1} \bar \w.
\end{aligned}
\end{equation}
Here, $\mathcal F(\omega, \omega')$ is the forcing term reported in Equation \eqref{eq:final_RF_forcing_term}.

\section{Connection to Dynamical Mean Field Theory}\label{sec:DMFT}

In \cite{bordelon2024dynamical}, it was shown how the SGD dynamics of linear regression and random feature models can be precisely characterized in the proportional asymptotic limit using methods of dynamical mean field theory (DMFT). The final answers are indeed in agreement with the test loss predictions derived in the prior sections. Further, in  \cite{bordelon2024dynamical}, the theoretical predictions---which we again emphasize coincide with those derived here---were tested extensively against numerics for both full batch gradient descent and SGD. Beyond just agreement of expressions, one also observes an interesting connection between the random matrix formalism presented in this work and dynamical mean field theory. 

The DMFT response functions ($\mathcal R_1, \mathcal R_3$ in the notation of \cite{bordelon2024dynamical}) correspond exactly to $1/S_{\W}$ and $1/S_{\F \F^\top}$ in this work, while the correlation functions $\mathcal C_0(\omega,\omega')$ in \cite{bordelon2024dynamical} correspond to our expressions for $\mathcal F(\omega,\omega')$ obtained with two-point deterministic equivalence. Indeed, under more general conditions, we find that the response functions arising in the DMFT treatment of a random matrix ensemble can be understood as $S$-transforms evaluated at $\df_1$ of an appropriate (renormalized) frequency. Further, the DMFT correlation functions are computing the two-point deterministic equivalents highlighted here.

\section{Conclusion}

We have derived a class of two-point deterministic equivalents of random matrices. Using this, we have been able to provide sharp asymptotics for the training, generalization, and out-of-distribution performance of a variety of linear models. Our results include both statics and dynamics. In all settings, we see that the $S$-transform of free probability plays a key role. Several of the results have been obtained in prior literature using either one-point equivalents in random matrix theory \cite{paquette20244plus} or via dynamical mean field theory \cite{bordelon2024dynamical}. Our approach provides a novel diagrammatic derivation of this two-frequency resolvent correlation that is key to capturing the non-commutative dynamics which arises in random feature models.  

\clearpage

\section*{Acknowledgements}

The authors are grateful to Bruno Loureiro, Hamza Chaudhry, Alex Wei, and Jamie Sully for discussions on random matrix theory. We also thank Benjamin Ruben for helpful comments on a previous version of this manuscript. 

AA and C. Pehlevan were supported by NSF Award DMS-2134157 and NSF CAREER Award IIS-2239780. B.B. is supported by a Google PhD Fellowship. JAZV is supported by the Office of the Director of the National Institutes of Health under Award Number DP5OD037354. The content is solely the responsibility of the authors and does not necessarily represent the official views of the National Institutes of Health. JAZV is further supported by a Junior Fellowship from the Harvard Society of Fellows. C. Pehlevan is further supported by a Sloan Research Fellowship. C. Paquette is a Canadian Institute for Advanced Research (CIFAR) AI chair, Quebec AI Institute (MILA) and a Sloan Research Fellow in Computer Science (2024). C. Paquette was supported by a Discovery Grant from the Natural Science and Engineering Research Council (NSERC) of Canada, NSERC CREATE grant Interdisciplinary Math and Artificial Intelligence Program (INTER-MATH-AI), Google research grant, and Fonds de recherche du Qu{\'e}bec - Nature et technologies (FRQNT) New University Researcher's Start-Up Program. This research is based on work supported by the CIFAR Pan-Canadian AI Strategy through a Catalyst award. Additional revenues related to this work: C. Paquette has 20\% part-time employment at Google DeepMind. This work has been made possible in part by a gift from the Chan Zuckerberg Initiative Foundation to establish the Kempner Institute for the Study of Natural and Artificial Intelligence.

\textbf{Author contributions:} AA, BB, and JAZ-V contributed equally to this work.

\bibliography{refs}

\appendix

\clearpage

\section{All Two-Point Deterministic Equivalences}\label{app:all_two_point}

In this section, we report all variants of the two-point deterministic equivalences. These extend prior equivalences observed in \cite{bach2024high,atanasov2024risk} to the case of different ridges $\lambda, \lambda'$. First, we have:
\begin{equation}\label{eq:master_2pt_1}
\begin{aligned}
    & (\lambda + \A \B)^{-1} \M (\lambda' + \A \B)^{-1} \simeq  S_{\B} S'_{\B} \left[ \G_{\A} \M \G'_{\A} +  \G_{\A} \A \G'_{\A} \frac{q \tr[\A \G_{\A} \M \G_{\A}']}{1 - q \df_2(\kappa, \kappa')} \right].
\end{aligned}
\end{equation}
One also obtains for $\A * \B = \A^{1/2} \B \A^{1/2}$ the same deterministic equivalence:
\begin{equation}\label{eq:master_2pt_12}
\begin{aligned}
    & (\lambda + \A *\B)^{-1} \M (\lambda' +  \A *\B)^{-1} \simeq  S_{\B} S'_{\B} \left[ \G_{\A} \M \G'_{\A} +  \G_{\A} \A \G'_{\A} \frac{q \tr[\A \G_{\A} \M \G_{\A}']}{1 - q \df_2(\kappa, \kappa')} \right].
\end{aligned}
\end{equation}
Additionally, one has:
\begin{equation}\label{eq:master_2pt_3}
\begin{aligned}    
    & (\lambda + \B \A )^{-1} \M (\lambda' + \A \B)^{-1} \simeq  S_{\B} S'_{\B} \left[ \G_{\A} \M \G'_{\A} +  \G_{\A} \G'_{\A} \frac{q \tr[\A^2 \G_{\A} \M \G_{\A}']}{1 - q \df_2(\kappa, \kappa')} \right].
\end{aligned}
\end{equation}
This was for resolvents. By the definition of $\T_{\A}, \T_{\A}'$, one immediately gets:
\begin{equation}\label{eq:master_2pt_2}
\begin{aligned}
    & \A \B (\lambda + \A \B)^{-1} \M  \A \B (\lambda' + \A \B)^{-1} \simeq \T_{\A} \M \T_{\A}' + \kappa \kappa' \G_{\A} \A \G_{\A}' \frac{q \tr[\A \G_{\A} \M \G_{\A}']}{1 - q \df_2(\kappa, \kappa')}.
\end{aligned}
\end{equation}
One has the same deterministic equivalence upon replacing $\A \B$ by the free product $\A * \B$.
\begin{equation}\label{eq:master_2pt_4}
\begin{aligned}
    & \A * \B (\lambda + \A * \B)^{-1} \M  \A * \B (\lambda' + \A * \B)^{-1} \simeq \T_{\A} \M \T_{\A}' + \kappa \kappa' \G_{\A} \A \G_{\A}' \frac{q \tr[\A \G_{\A} \M \G_{\A}']}{1 - q \df_2(\kappa, \kappa')}.
\end{aligned}
\end{equation}
Again, for Gram matrices, the above equations hold but with $q \to 1/q$.

By adopting the notation $\hat{\G}_{\A} = \G_{\A \B}$, $\hat{\T}_{\A} = \T_{\A \B}$, $\gamma = q \df_2(\kappa, \kappa')$ and $\gamma_{\M} =q  \tr[\A \G_{\A} \M \G_{\A}] $ one can write these equivalences as:
\begin{align}
    \hat \G_{\A} \M \hat \G_{\A}' &\simeq S S' \G_{\A} \M \G_{\A}' + S S' \G_{\A}\A \G_{\A}' \frac{\gamma_{\M}}{1-\gamma},  \\
    \hat \T_{\A} \M \hat \T_{\A}' &\simeq  \T_{\A} \M \T_{\A}' + \kappa \kappa' \G_{\A}\A \G_{\A}' \frac{\gamma_{\M}}{1-\gamma}, \\
    \hat \G_{\A} \M \hat \T_{\A}' &\simeq  S \G_{\A} \M \T_{\A}' - S \kappa'  \G_{\A}\A \G_{\A}' \frac{\gamma_{\M}}{1-\gamma}, \\
    \hat \T_{\A} \M \hat \G_{\A}' &\simeq  S' \T_{\A} \M \G_{\A}' - \kappa S' \G_{\A}\A \G_{\A}' \frac{\gamma_{\M}}{1-\gamma}.
\end{align}

\subsection{Sanity Check of Two Point Functions}

We now take $\A = \S, \B = \W$ for $\W$ a white Wishart and consider the matrix $\Sh = \S * \W$. In the case where $\M = \mathbf I$, the last two deterministic equivalences are equal. Further, by taking $\lambda = \lambda'$ so that $\kappa = \kappa'$ we get:
\begin{equation}
\begin{aligned}
    \Sh (\Sh + \lambda)^{-2} &\simeq S \S (\S+ \kappa)^{-2} - S \kappa   \S (\S+ \kappa)^{-2} \frac{q \tr[\S (\S+\kappa)^{-2}] }{1 - \gamma}\\
    &= \S (\S+ \kappa)^{-2} S \left[ 1 - \frac{ - q \kappa \df_1'}{1-\gamma} \right]\\
    &= \frac{1}{1-\gamma} \S (\S+ \kappa)^{-2} S \left[ 1- \gamma + q (\df_1 - \df_2) \right]\\
    &= \frac{1}{1-\gamma} \S (\S+ \kappa)^{-2}\\
    &= \frac{d \kappa}{d \lambda} \S (\S+ \kappa)^{-2}.
\end{aligned}
\end{equation}
So we see that the the two-point result yields the same result as would be obtained through differentiation. This also extends to $\lambda, \lambda'$ not equal. Taking $\df_2(\lambda, \lambda') = \tr[\S^2 (\S +\lambda)^{-1} (\S+ \lambda')^{-1}]$ and $\gamma = \frac{D}{P} \df_2(\lambda, \lambda')$, we have:
\begin{equation}\label{eq:mixed_2_pt_simplify}
\begin{aligned}
    &\Sh (\Sh + \lambda)^{-1} (\Sh +\lambda')^{-1} \\&\quad\simeq  \S (\S + \kappa)^{-1} (\S +\kappa')^{-1} S \left(1 - 
 \kappa' \frac{q \tr[\S (\S + \kappa)^{-1} (\S + \kappa')^{-1}]}{1 - \gamma} \right)\\
 &\quad= \frac{1}{1-\gamma} \S (\S + \kappa)^{-1} (\S +\kappa')^{-1} S (1 - \gamma + q (\df_1 - \df_2))\\
 &\quad= \frac{1}{1-\gamma}  \S (\S + \kappa)^{-1} (\S +\kappa')^{-1}.
\end{aligned}
\end{equation}
Here we have used that $\gamma = q \df_2$ and $S = (1-q \df_1)^{-1}$.
This equation will be useful in treating various SGD kernel-related quantities. 

\section{Double-Frequency Equivalents for the Empirical Forcing and SGD Kernel Terms}

In this section we match the double-frequency treatment of the population forcing term. For the empirical forcing terms, the bi-frequency picture yields an analogue of the generalized cross-validation (GCV) procedure in each frequency mode. The utility of the bi-frequency picture for the kernel forcing terms is less clear, but we report it anyway for completeness.

\subsection{Linear Regression Forcing Term}\label{app:double_frequency_forcing}

We can write the empirical loss under gradient flow in terms of two time variables as:
\begin{equation}
    \hat{\mathcal F}(t, t') \equiv \Delta \w_t^\top \Sh \Delta \w_{t'} = \bar \w^\top e^{-t \Sh} \Sh e^{-t' \Sh} \bar \w.
\end{equation}
Then, Fourier transforming in each variable separately and restricting to the $t = t'$ diagonal yields:
\begin{equation}
    \hat{\mathcal F}(t) = \int_{\omega,\omega'} e^{i t (\omega + \omega')} \underbrace{\bar \w^\top \Sh (\Sh + i \omega)^{-1} (\Sh + i \omega')^{-1} \bar \w}_{\hat{\mathcal F}(\omega, \omega')} .
\end{equation}
We now apply the deterministic equivalence \eqref{eq:mixed_2_pt_simplify} to obtain:
\begin{equation}\label{eq:LR_empirical_forcing_det_equiv}
    \hat{\mathcal F}(\omega, \omega') \simeq  \frac{1}{1 - \gamma}  \bar \w^\top  \S (\S + i \omega_1)^{-1}  (\S + i \omega_1')^{-1} \bar \w.
\end{equation}

\subsection{SGD in Linear Regression}\label{app:double_frequency_kernel}

Here, to match the gradient flow term, we apply two Fourier transforms to the kernel and apply the two-point deterministic equivalence. Namely, we consider extending $\mathcal K, \hat {\mathcal K}$ to:
\begin{equation}
    \mathcal K(t, t') \equiv \Tr[e^{-t \Sh} \Sh e^{-t' \Sh} \S], \quad \hat{\mathcal K}(t, t') \equiv \Tr[e^{-t \Sh} \Sh e^{-t' \Sh} \Sh].
\end{equation}
We then perform Fourier transforms separately in $t, t'$ to obtain:
\begin{equation}
\begin{aligned}
     \mathcal K(t) &= D \int_{\omega,\omega'} e^{i t (\omega + \omega') } \underbrace{\tr[\S (\Sh + i \omega)^{-1} \Sh (\Sh + i \omega')^{-1} ]}_{\mathcal K(\omega, \omega')} , \\
      \hat{\mathcal K}(t) &=  D \int_{\omega,\omega'} e^{i t (\omega + \omega') } \underbrace{ \tr[\Sh (\Sh + i \omega)^{-1} \Sh (\Sh + i \omega')^{-1} ] }_{\hat{\mathcal K}(\omega, \omega')} .
\end{aligned}
\end{equation}
By applying the two-point master formulas in Appendix \ref{app:all_two_point}, and specifically \eqref{eq:mixed_2_pt_simplify}, we obtain the equivalent for the test kernel:
\begin{equation}
\begin{aligned}
    \mathcal K(\omega, \omega') &\simeq  \frac{D \df_2 }{1-\gamma}.
\end{aligned}
\end{equation}
Similarly the train kernel can be written as:
\begin{equation}
\begin{aligned}
    \hat{\mathcal K}(\omega, \omega') &\simeq D \df_2 \, \left( 1 - (i \omega_1) (i \omega_1') \frac{D}{P} \frac{ \tr[\S (\S+ i \omega_1)^{-1} (\S + i \omega_1')^{-1}]^2 }{1 - \gamma} \right).
\end{aligned}
\end{equation}

\subsection{Linear Random Features Forcing Term}\label{sec:RF_GCV}

The empirical loss under gradient flow in bi-frequency space is given by
\begin{equation}
    \hat{\mathcal F}(t) = \int_{\omega,\omega'} e^{i t (\omega + \omega')} \bar \w^\top \Sh (\F \F^\top \Sh + i \omega)^{-1} (\F \F^\top \Sh + i \omega')^{-1} \bar \w.
\end{equation}
We now apply the deterministic equivalence \eqref{eq:mixed_2_pt_simplify} to obtain:
\begin{equation}
    \hat{\mathcal F}(\omega, \omega') \simeq  \frac{1}{1 - \gamma_1}  \bar \w^\top  \S (\S \F \F^\top + i \omega_1)^{-1}  (\S \F \F^\top + i \omega_1')^{-1} \bar \w.
\end{equation}
Finally, we apply the deterministic equivalence \eqref{eq:master_2pt_1} over $\F \F^\top$ to yield:
\begin{equation}
\begin{aligned}
    \hat{\mathcal F}(\omega, \omega') &\simeq  \frac{S_{\F \F^\top} S_{\F \F^\top}'}{1 - \gamma_1} \Bigg[ \bar \w^\top  \S (\S+ i \omega_2)^{-1}  (\S + i \omega_2')^{-1} \bar \w \\
    & \qquad\qquad\qquad\quad + \bar \w^\top (i \omega_2 + \S)^{-1} \S^2 (i \omega'_2 +  \S)^{-1} \bar \w  \frac{\frac{D}{N} \tr[ (i \omega_2 + \S)^{-1} \S (i \omega'_2 +  \S)^{-1}]}{1 - \frac{D}{N} \df_2(i \omega_2, i \omega_2')} \Bigg]\\
    &= \frac{\mathcal F(\omega, \omega')}{S_{\W} S_{\W}'}.
\end{aligned}
\end{equation}
This again yields a dynamical analogue of GCV, where we see that the empirical and population risks under gradient flow differ by a factor of $S_{\W} S_{\W}'$. In the $t \to \infty$ limit this requires the $S_{\W}^2$ obtained in \cite{adlam2020neural, atanasov2024scaling} for linear random features.

\subsection{SGD in Linear Random Features}

We now compute deterministic equivalents for the kernel term, again defining the bi-temporal kernels by:
\begin{equation}
\begin{aligned}
    \mathcal K(t, t') &\equiv \Tr[e^{-t \F \F^\top \Sh } \F\F^\top \Sh  e^{-t' \F \F^\top \Sh} \F \F^\top \S], \\
    \hat{\mathcal K}(t, t') &\equiv \Tr[e^{-t \F \F^\top \Sh} \F \F^\top \Sh e^{-t' \F \F^\top \Sh} \F \F^\top \Sh].
\end{aligned}
\end{equation}
We then perform Fourier transforms separately in $t, t'$ to obtain:
\begin{equation}
\begin{aligned}
     \mathcal K(t) &=D \int_{\omega,\omega'} e^{i t (\omega + \omega') } \mathcal K(\omega, \omega') , \\
      \hat{\mathcal K}(t) &= D \int_{\omega,\omega'} e^{i t (\omega + \omega') } \hat{\mathcal K}(\omega, \omega')
\end{aligned}
\end{equation}
for
\begin{equation}
\begin{aligned}
    \mathcal K(\omega, \omega') &\equiv \tr[\F \F^\top \S (\F \F^\top \Sh + i \omega)^{-1} \F \F^\top \Sh ( \F \F^\top \Sh + i \omega')^{-1} ]
    \\ 
    \hat{\mathcal K}(\omega, \omega') &\equiv \tr[\F \F^\top \Sh ( \F \F^\top \Sh + i \omega)^{-1} \F \F^\top \Sh (\F \F^\top \Sh + i \omega')^{-1} ] . 
\end{aligned}
\end{equation}
Applying the two-point equivalences to perform the data average, we get for the test kernel:
\begin{equation}
    \mathcal K(\omega, \omega') \simeq  \frac{D \df_{\F\F^\top \S}^2 (i \omega_1, i \omega_1')}{1-\gamma_1(i \omega_1, i \omega_1')}.
\end{equation}
We recognize again that the numerator and denominator depend only on $\gamma_1 \equiv \frac{D}{P} \df^2_{\F\F^\top \S}$, We have already computed equivalents for this in the prior section. 

Similarly for the train kernel we get:
\begin{equation}
\begin{aligned}
    \hat{\mathcal K}(\omega, \omega') &\simeq D \df^2_{\F \F^\top \S}(i \omega_1, i \omega_1')  \left[ 1 - (i \omega_1) (i \omega_1') \frac{D}{P} \frac{ \tr[\S \F \F^\top (\S \F \F^\top\!\! + i \omega_1)^{-1} (\S \F \F^\top\!\! + i \omega_1')^{-1}] }{1 - \gamma_{1}(i \omega_1, i \omega_1')} \right].
\end{aligned}
\end{equation}
Again equivalents for $\df^2_{\F \F^\top \S}$ and $\gamma_1$ are already calculated. It remains to apply a final deterministic equivalence  \eqref{eq:mixed_2_pt_simplify}  over the features to the last term to get:
\begin{equation}
\begin{aligned}
    &\tr[\S \F \F^\top (\S \F \F^\top\!\! + i \omega_1)^{-1} (\S \F \F^\top\!\! + i \omega_1')^{-1}]  \simeq \frac{\tr[\S  (\S + i \omega_2)^{-1} (\S + i \omega_2')^{-1}]}{1 - \frac{D}{N} \df_2(i \omega_2, i \omega_2')}.
\end{aligned}
\end{equation}

\section{Covariate Shift in Random Features}\label{app:RF_OOD}

In this section we write down the exact formula for the precise asymptotics of random feature regression when tested under covariate shift. These formulas are obtained from a straightforward though slightly tedious application of the two-point deterministic equivalences derived in the text. 
\subsection{Gradient Flow Term}

We are interested in evaluating the test error out-of-distribution. Under gradient flow this can be written as:
\begin{equation}
    \mathcal{F}(i \omega, i \omega') = \bar \w^\top (i \omega + \Sh  \F \F^\top )^{-1}  \S' \F \F^\top  (i \omega' +   \Sh \F \F^\top )^{-1} \,  (\F \F^\top)^{-1} \bar \w.
\end{equation}

\subsubsection{Data Average}

We apply \eqref{eq:master_2pt_1} with $\M = \A =\S  \F \F^\top $ to obtain.
\begin{equation}
\begin{aligned}
    &(i \omega + \Sh \F \F^\top  )^{-1} \S \F \F^\top (i \omega' +  \Sh  \F \F^\top)^{-1}\\
    &\quad \simeq S_{\W} S_{\W}' (i \omega_1 + \S \F \F^\top )^{-1} \S' \F \F^\top  (i \omega_1' + \S \F \F^\top )^{-1} \\
    & \qquad + S_{\W} S_{\W} \frac{\gamma_1'}{1-\gamma_1}  (i \omega_1 + \S \F \F^\top )^{-1} \S \F \F^\top  (i \omega_1' + \S \F \F^\top )^{-1}.
\end{aligned}
\end{equation}
Here, we have
\begin{equation}
\begin{aligned}
    \gamma_1 &= \frac{D}{P} \tr\left[ ( \S \F \F^\top )^{2} (i \omega_1 + \S \F \F^\top)^{-1} (i \omega_1' +  \S \F \F^\top)^{-1} \right],\\
    \gamma_1' &= \frac{D}{P} \tr\left[  \S' \F \F^\top  \S \F \F^\top  (i \omega_1 + \S \F \F^\top)^{-1} (i \omega_1' +  \S \F \F^\top)^{-1} \right].
\end{aligned}
\end{equation}
\subsubsection{Feature Average}

We have already evaluated $\gamma_1$ in the main text. $\gamma_1'$ is similarly straightforward and is evaluated to be:
\begin{equation}
\begin{aligned}
        \gamma_1' &= \frac{D}{P} \df^2_{\S,\S'}(i \omega_2, i \omega_2') \\
        & \quad + \frac{D}{P} (i \omega_2)(i \omega'_2) \frac{ \frac{D}{N}  }{1 - \frac{D}{N} \df_2(i \omega_2, i \omega'_2) }
        \tr[ (i \omega_2 + \S)^{-1} \S' (i \omega'_2 +  \S)^{-1} ] \tr[ (i \omega_2 + \S)^{-1} \S (i \omega'_2 +  \S)^{-1} ] . 
\end{aligned}
\end{equation}
The remaining part of the forcing term that has not yet been evaluated can gain be rewritten via the push-through identity as: 
\begin{equation}
\begin{aligned}
    & (i \omega_1 + \S \F \F^\top )^{-1} \S' \F \F^\top  (i \omega_1' + \S \F \F^\top )^{-1} (\F \F^\top)^{-1}  =  (i \omega_1 + \S \F \F^\top )^{-1} \S'  (i \omega_1' + \F \F^\top  \S )^{-1}.
\end{aligned}
\end{equation}
We now apply  \eqref{eq:master_2pt_0} with $\M = \mathbf I, \A = \S, \B = \F \F^\top$. After some rewriting, we obtain:
\begin{equation}
\begin{aligned}
     \mathcal F_{OOD}(i \omega, i \omega') =  \gamma_1' \mathcal F(i \omega, i \omega') + S S' \Bigg[ & \bar \w^\top (i \omega_2 + \S)^{-1} \S' (i \omega'_2 +  \S)^{-1} \bar \w  \\
     & + \bar \w^\top (i \omega_2 + \S)^{-1} \S^2 (i \omega'_2 +  \S)^{-1} \bar \w  \frac{\frac{D}{N} \tr[ (i \omega_2 + \S)^{-1} \S' (i \omega'_2 +  \S)^{-1}]}{1 - \frac{D}{N} \df_2(i \omega_2, i \omega_2')} \Bigg].
\end{aligned}
\end{equation}
Here, as before, $\mathcal F(i \omega, i \omega')$ is the in-distribution test error.

\subsection{SGD Kernel Term}
By applying a one-point deterministic equivalence, the OOD test kernel is:
\begin{equation}
\begin{aligned}
    \mathcal K_{OOD}(\omega, \omega') &\simeq  \Tr[\S' \S(\S + i \omega_2)^{-1}] + \frac{i \omega_2}{N} \Tr[\S' (\S + i \omega_2)^{-1}] \Tr[\S (\S + i \omega_2)^{-1}].
\end{aligned}
\end{equation}
The train kernel is (by definition) unaffected by distribution shift.

\end{document}